\documentclass[aps,prd,nofootinbib,floatfix,amsmath,amssymb,twocolumn]{revtex4}
\usepackage{mathrsfs}
\usepackage{graphicx}
\usepackage[dvipsnames]{xcolor}
\usepackage{dsfont}
\usepackage{tikz}
\begin{document}

\makeatletter
%Feynman slash
\newbox\slashbox \setbox\slashbox=\hbox{$/$}
\newbox\Slashbox \setbox\Slashbox=\hbox{\large$/$}
\def\pFMslash#1{\setbox\@tempboxa=\hbox{$#1$}
  \@tempdima=0.5\wd\slashbox \advance\@tempdima 0.5\wd\@tempboxa
  \copy\slashbox \kern-\@tempdima \box\@tempboxa}
\def\pFMSlash#1{\setbox\@tempboxa=\hbox{$#1$}
  \@tempdima=0.5\wd\Slashbox \advance\@tempdima 0.5\wd\@tempboxa
  \copy\Slashbox \kern-\@tempdima \box\@tempboxa}
\def\FMslash{\protect\pFMslash}
\def\FMSlash{\protect\pFMSlash}
\def\miss#1{\ifmmode{/\mkern-11mu #1}\else{${/\mkern-11mu #1}$}\fi}
%%%% Uso:  \pFMSlash{p}
\makeatother

%\tightenlines
\title{Charged lepton flavor violating decays $Z\to \ell_\alpha\ell_\beta$ in the inverse seesaw}

\author{Adri\'an Gonz\'alez-Quiterio and H\'ector Novales-S\'anchez}
\affiliation{
Facultad de Ciencias F\'isico Matem\'aticas, Benem\'erita Universidad Aut\'onoma de Puebla, Apartado Postal 1152 Puebla, Puebla, M\'exico}

\begin{abstract}
After confirmation of the massiveness and mixing of neutrinos, by neutrino oscillation data, the origin of neutrino mass and the occurrence of charged-lepton-flavor non-conservation in nature have become two main objectives for the physics of elementary particles. Taking inspiration from these  two matters, we address the $Z$-boson lepton decays $Z\to\ell_\alpha\ell_\beta$, with $\ell_\alpha\ne\ell_\beta$, thus violating charged-lepton flavor. We calculate the set of contributing one-loop Feynman diagrams characterized by virtual neutral leptons, both light and heavy, emerged from the inverse seesaw mechanism for the generation of neutrino mass. \textcolor{Black}{We obtain general analytical results, which decouple in the limit as $\Lambda\to\infty$, with $\Lambda$ the high-energy scale involved in the inverse seesaw mechanism. Our quantitative analyses consider two scenarios of heavy-neutral-lepton mass: (1) the mass spectrum is degenerate; and (2) a more complete framework which includes both degenerate and non-degenerate mass spectra. The quantitative estimations we perform in both scenarios take into account upper bounds on non-unitarity deviations in the lepton mixing matrix and perturbativity of Yukawa couplings.} We find that, even though current constraints on $Z\to\ell_\alpha\ell_\beta$, by the ATLAS Collaboration, remain far from inverse-seesaw contributions, improved sensitivity from in-plans machines, such as the Future Circular Collider and the Circular Electron Positron Collider, shall be able to probe this mass-generating mechanism through \textcolor{Black}{the decay $Z\to\tau e$, whose contributions could be as large as $10^{-7}$.}
\end{abstract}


\maketitle

\section{Introduction}
\label{intro}
Nowadays, the neutrino sector is, doubtless, one of the most appealing topics in particle physics. The experimental confirmation of neutrino oscillations~\cite{Pontecorvo}, first reported by the Super Kamiokande~\cite{SKamiokande} and the Sudbury Neutrino Observatory~\cite{SNO}, showed that there is, indeed, unknown physics beyond our beloved Standard Model (SM)~\cite{Glashow,Weinberg,Salam}. Since then, a huge amount of work has been devoted to understand the physics underlying neutrinos, aiming at the elucidation of a number of aspects, such as the one mechanism behind neutrino-mass generation~\cite{AsSh,ABS,PaSa,MoSe1,MoSe2,Pilaftsis,MoVa,GoVa,DeVa,ALSV1,ALSV2}, whether their description corresponds to the Dirac~\cite{Dirac} or the Majorana~\cite{Majorana} theories, and the possible link among neutrinos and dark matter~\cite{DoWi,ShFu,Kusenko,Ma}. Regarding the mechanism for neutrino-mass generation, several proposals are available, the most economical consisting in endowing neutrinos with masses in the same way as it is done with the rest of the fermions of the SM, namely, by the inclusion of Yukawa terms $\overline{L_L}Y_\nu\tilde{\phi}\nu_R+\textrm{H.c.}$, featuring right-handed neutrino fields, $\nu_R$, assumed to be singlets under the electroweak SM gauge group ${\rm SU}(2)_L\otimes{\rm U}(1)_Y$. Nonetheless, due to current upper bounds on neutrino mass~\cite{KATRIN,eBOSS,PlanckCollab,DESICollab,CUORE,GERDA,KamLANDZen}, this framework requires the assumption of tiny Yukawa couplings, of order $\sim10^{-12}$, which, though not forbidden, can be avoided in alternative schemes, coming along with further interesting possibilities of new physics. In particular, the inclusion of renormalizable Majorana-mass like terms $\frac{1}{2}\overline{\nu^{\rm c}_R}m_{\rm M}\nu_R+{\rm H.c.}$, with $m_{\rm M}$ assumed to originate from some high-energy description beyond SM and where $\psi^{\rm c}=C\overline{\psi}^{\rm T}$ is the charge-conjugate field of $\psi$, yields the neutrino mass matrix
\begin{equation}
M_\nu
=
\left(
\begin{array}{cc}
0 & m_{\rm D}
\vspace{0.2cm}
\\
m_{\rm D}^{\rm T} & m_{\rm M}
\end{array}
\right),
\label{T1seesaw}
\end{equation}
where $m_{\rm D}\propto v$, with $v=246\,\textrm{GeV}$ the vacuum expectation value of the SM Higgs doublet. Moreover, $m_\textrm{M}\propto\Lambda$, with $\Lambda$ the high-energy scale associated to the new-physics description. After a Takagi diagonalization~\cite{Takagi} and under the assumption of a very large high-energy scale $\Lambda$, the matrix $M_\nu$, Eq.~\eqref{T1seesaw}, gives rise to mass terms for light neutrinos, $n_j$, and heavy neutrinos, $N_j$, both of Majorana type, abiding by the mass profile $m_{n_j}\sim\frac{v^2}{\Lambda}$ and $m_{N_j}\sim\Lambda$. This is the well-known ``seesaw mechanism''~\cite{PaSa,MoSe1,MoSe2} for the generation of neutrino mass, or, more precisely, its type-1 variant. Even though the seesaw mechanism provides an elegant mean to explain the noticeable smallness of the masses of the known neutrinos, from here on dubbed ``light neutrinos'', it bears a practical drawback: such small neutrino masses imply that the scale $\Lambda$ must be enormous, of order $\sim10^{13}\,\textrm{GeV}$, thus preventing direct production of heavy neutrinos, by current and even future experimental facilities, and also imposing a large suppression on virtual effects from the heavy neutrinos. In short, such a scheme is quite difficult to probe. 
\\

In order to bring new physics closer to a possible measurement, much theoretical work has been devoted to model building. Therefore, a large number of seesaw variants, for neutrino-mass generation, has been propounded\footnote{Refs.~\cite{CHSVV,CHLR} provide nice reviews on seesaw-type mechanisms and more.}. In this context, the so-called ``inverse seesaw mechanism'' (ISSM)~\cite{MoVa,GoVa,DeVa}, around which the investigation detailed in the present work has been developed, has attracted much attention. For this mechanism to operate, three right-handed fermion fields, denoted by $\nu_{j,R}$ with $j=1,2,3$, augment the field content of the SM. Then, three further left-handed fermion fields $S_{j,L}$, where $j=1,2,3$, are also added. All these fermion fields are assumed to be singlets under the SM gauge group, with lepton-number assignments $L(\nu_R)=+1$ and $L(S_L)=+1$. After breaking of the electroweak gauge symmetry group into the electromagnetic group, the neutrino-mass Lagrangian turns out to be given by
\begin{eqnarray}
&&
{\cal L}_{\nu}=-\overline{\nu_L}m_\textrm{D}\nu_R-\overline{S_L}M\nu_R
\nonumber \\ && \hspace{0.8cm}
-\frac{1}{2}\overline{S_L}\mu_SS^{\rm c}_L-\frac{1}{2}\overline{\nu_R^{\rm c}}\mu_R\nu_R+\textrm{H.c.}
\label{renmassterms}
\end{eqnarray}
The last two terms of the right-hand side of this equation break the global symmetry $\textrm{U}(1)$. In such case, the matrices $\mu_S$ and $\mu_R$ can be assumed to be small on the grounds of naturalness, in the sense of t'Hooft~\cite{naturalness}. Moreover, a connection among $\mu_S$, $\mu_R$ and some low-energy scale $v_\sigma$, connected to the breaking of the global symmetry with respect to $\textrm{U}(1)$, is assumed, namely, $\mu_S\propto v_\sigma$ and $\mu_R\propto v_\sigma$. Bearing the previous discussion in mind, from here on we assume the energy-scales hierarchy
\begin{equation}
\Lambda\gg v\gg v_\sigma.
\label{hierarchy}
\end{equation}
Once the breaking of the electroweak gauge symmetry has taken place, the definitions
\begin{equation}
M_{\rm M}
=
\left(
\begin{array}{cc}
\mu_R & M^{\rm T}
\vspace{0.2cm}
\\
M & \mu_S
\end{array}
\right),
\end{equation}
%%%%%%%%%%
\begin{equation}
M_{\rm D}
=
\left(
\begin{array}{cc}
m_{\rm D} & 0
\end{array}
\right),
\end{equation}
are used, which allow one to rearrange the neutrino-mass Lagrangian $\mathcal{L}_\nu$ in terms of a neutrino-mass matrix with the same structure as the one shown in Eq.~\eqref{T1seesaw}, but with the replacements $m_{\rm D}\to M_{\rm D}$ and $m_{\rm M}\to M_{\rm M}$. Such a neutrino-mass matrix is symmetric, so there exists a $9\times9$ unitary matrix, $\Omega$, which diagonalizes it as~\cite{Takagi}
\begin{equation}
M_\psi=\Omega^\textrm{T}\left(
\begin{array}{cc}
0 & M_{\rm D}
\vspace{0.2cm}
\\
M_{\rm D}^{\rm T} & M_{\rm M}
\end{array}
\right)\Omega
=
\left(
\begin{array}{ccc}
M_n & 0 & 0
\vspace{0.2cm}
\\
0 & M_N & 0
\vspace{0.2cm}
\\
0 & 0 & M_X
\end{array}
\right),
\label{numassmatrix}
\end{equation}
where $M_n$, $M_N$, $M_X$ are $3\times3$ sized, diagonal, and real, with positive eigenvalues corresponding to: (1) the masses of the three light neutrinos $n_j$, in the case of $M_n$; (2) to the masses of three heavy neutrinos $N_j$, in the case of $M_N$; and (3) to the masses of three further heavy neutral leptons $X_j$, in the case of $M_X$. From here on, we indistinctly use the name ``heavy neutral leptons'' (HNLs) to refer to fermion fields $N_j$ or $X_j$. \textcolor{Black}{As discussed in Refs.~\cite{AHMW,HIY,GoNo}, the HNL mass matrices can be written as
\begin{eqnarray}
M_N=M-\frac{1}{2}\big( \mu_S+\mu_R \big),
\label{MNmatrix}
%\end{equation}
%%%%%%%%%%
%\begin{equation}
\\
M_X=M+\frac{1}{2}\big( \mu_S+\mu_R \big),
\label{MXmatrix}
\end{eqnarray}
\\
so they differ by $M_X-M_N=\mu_R+\mu_S\sim v_\sigma$, thus implying that} the mass spectrum of the HNLs $N_j$ and $X_j$ is quasi-degenerate by pairs, that is, $m_{N_j}\approx m_{X_j}$, for $j=1,2,3$, though keep in mind that, in general, $m_{N_j}\ne m_{N_k}$ if $j\ne k$, so a full near-degenerate mass spectrum, in which the 6 HNLs almost share the same mass value, can be assumed, but it does not hold in general. The unitary diagonalization matrix $\Omega$ can be expressed as the product 
\begin{equation}
\Omega=UV, 
\label{omegaUV}
\end{equation}
of two $9\times9$ unitary matrices $U$ and $V$. Since $\Lambda\gg v$, as established in Eq.~\eqref{hierarchy}, all relevant quantities in the model are expressed in terms of some order of the matrix product $M_{\rm D}M_{\rm M}^{-1}$. In this context, the unitary matrix $U$ can be expressed as
\begin{widetext}
\begin{equation}
U=\left(
\begin{array}{cc}
\big(\textbf{1}_3+(M_{\rm D}M_{\rm M}^{-1})^*(M_{\rm D}M_{\rm M}^{-1})^\textrm{T}\big)^{-\frac{1}{2}} & (M_{\rm D}M_{\rm M}^{-1})^*\big(\textbf{1}_6+(M_{\rm D}M_{\rm M}^{-1})^\textrm{T}(M_{\rm D}M_{\rm M}^{-1})^*\big)^{-\frac{1}{2}}
\vspace{0.5cm} \\
-(M_{\rm D}M_{\rm M}^{-1})^\textrm{T}\big(\textbf{1}_3+(M_{\rm D}M_{\rm M}^{-1})^*(M_{\rm D}M_{\rm M}^{-1})^\textrm{T}\big)^{-\frac{1}{2}} & \big(\textbf{1}_6+(M_{\rm D}M_{\rm M}^{-1})^\textrm{T}(M_{\rm D}M_{\rm M}^{-1})^*\big)^{-\frac{1}{2}}
\end{array}
\right),
\end{equation}
%\begin{equation}
%U
%\approx
%\left(
%\begin{array}{cc}
%{\bf 1}_3-\frac{1}{2}M_{\rm D}^*\big( M_{\rm M}^{-1} \big)^*M_{\rm M}^{-1}M_{\rm D}^{\rm T} & M_{\rm D}^*\big( M_{\rm M}^{-1} \big)^*
%\vspace{0.2cm}
%\\
%-M_{\rm M}^{-1}M_{\rm D}^{\rm T} & {\bf 1}_6-\frac{1}{2}M_{\rm M}^{-1}M_{\rm D}^{\rm T}M_{\rm D}^*\big( M_{\rm M}^{-1} \big)^*
%\end{array}
%\right),
%\label{Ublockpar}
%\end{equation}
\end{widetext}
where ${\bf 1}_n$ denotes the identity matrix of size $n\times n$. Let us remark that the role of the matrix $U$ is to block-diagonalize the neutrino-mass matrix. The remaining diagonalizations are carried out by the $9\times9$ matrix $V$, which is expressed as
\begin{equation}
V
=
\left(
\begin{array}{cc}
U^*_{\rm PMNS} & 0
\vspace{0.2cm}
\\
0 & \tilde{V}
\end{array}
\label{Vmatrix}
\right),
\end{equation}
with $U_{\rm PMNS}$ the lepton-mixing matrix, also known as the Pontecorvo-Maki-Nakagawa-Sakata (PMNS) matrix~\cite{MNSinPMNS,PinPMNS}, and where $\tilde{V}$ is some $6\times6$ unitary matrix, aimed at the diagonalization of the HNL mass matrices. In this context, the emblematic ISSM relation for the light-neutrino masses,
\begin{equation}
M_n\approx U_{\rm PMNS}^\dag m_{\rm D}M^{-1}\mu_S\big( M^{\rm T} \big)^{-1}m_{\rm D}^{\rm T}U_{\rm PMNS}^*,
\label{issrelation}
\end{equation}
emerges. An interesting aspect regarding Eq.~\eqref{issrelation} is the role played by the matrix $\mu_S\sim v_\sigma$, since the smallness of this factor reduces the pressure on the high-energy scale $\Lambda$, appearing in $M^{-1}\sim\frac{1}{\Lambda}$, thus allowing for much smaller values of this scale, in comparison with the situation that takes place in the type-1 seesaw mechanism, thus bringing the effects of the new physics closer. Moreover, since $M_N,M_X \approx M\sim \Lambda$, this also reduces, in passing, the sizes of the masses of the six HNLs $N_j$ and $X_j$.
\\

Charged-lepton-flavor-violation (cLFV) processes are forbidden in the SM. The occurrence of neutrino oscillations in nature, with the implication that neutrinos are massive and mix~\cite{MNSinPMNS}, means, among other things, that physical processes in which lepton flavor is not preserved are allowed, so two relevant questions which straightforwardly follow regard how large the contributions to such processes are and whether they can be sensed by experimental facilities. For instance, in the minimal extension of the SM in which the three light neutrinos receive masses from Dirac-Yukawa terms the cLFV decays $\ell_\alpha\to\gamma\,\ell_\beta$, with $\ell_\alpha$ and $\ell_\beta$ respectively denoting $\alpha$- and $\beta$-flavored charged leptons, are dramatically suppressed as an outcome of the Glashow-Iliopoulos-Maiani (GIM) mechanism~\cite{GIM}, with branching ratios as tiny as $\sim10^{-54}$, therefore being well out of the reach of experimental sensitivity, even in the long term. In the presence of HNLs, the aforementioned mechanism does not necessarily impose such a suppression, thus giving rise to much larger branching ratios, in some cases within the reach of current experimental sensitivity. This has been discussed, for instance, in Ref.~\cite{RNVS}. Another sort of processes in which lepton flavor is not conserved comprises the $Z$ boson decays into lepton pairs of different flavors, that is, $Z\to\ell_\alpha\ell_\beta$, with $\ell_\alpha\ne\ell_\beta$. In the present investigation, a calculation, at the one-loop level, of the contributions from neutral leptons, light ones as well as heavy ones, to the cLFV processes $Z\to\ell_\alpha\ell_\beta$ is performed, in the context of the ISSM for neutrino-mass generation. The resulting expressions are then used to carry out a quantitative analysis in which our estimations are compared with current experimental bounds on these $Z$-boson decays, recently established by the ATLAS Collaboration~\cite{ATLASZtotaualgo,ATLASZtoemu}, at the Large Hadron Collider (LHC). We also analyze our results considering projections on experimental sensitivity of the Future Circular Collider in its electron-positron phase (FCC-ee) and the Circular Electron-Positron Collider (CEPC) to these decay processes~\cite{FCCeebounds,CEPCbounds}. \textcolor{Black}{For our quantitative estimations, we utilize bounds on non-unitarity effects emerging in light-lepton mixing, reported in Refs.~\cite{BFHLMN,GoNo}, which serve as a further restriction, in addition to the ATLAS results. We also take into account the constraints imposed by keeping the Yukawa coupling $m_\textrm{D}$ within the perturbative regime. We find that the contributions from inverse-seesaw neutral leptons are well beyond current experimental sensitivity, by about $\sim2$ orders of magnitude in the most promising channel, which turns out to be $Z\to\tau e$. However, we note that the FCC-ee and CEPC, both of them in-plans facilities, would be able to probe the ISSM by means of this decay process. ISSM contributions to this branching ratio are found to be as large as $\textrm{Br}\big( Z\to\tau e \big)\sim10^{-7}$.}
\\

\textcolor{Black}{It is worth commenting that the cLFV decays $Z\to\ell_\alpha\ell_\beta$ have been addressed before, in the same theoretical framework, in Refs.~\cite{ARMOT,RHMS}. To this respect, we would like to remark that these studies work under the assumption that the mass spectrum of the HNLs is degenerate. Recall, however, that while the relation $m_{N_j}\approx m_{X_j}$, for $j=1,2,3$, holds, in general $m_{N_j}\ne m_{X_k}$, for $j\ne k$. In the present work we explored the possibility of having non-degenerate mass spectra of HNLs, aiming at quantitative estimations and at a comparison with the degenerate-mass scenario. In this sense, our work is comprehensive and complements the aforementioned references\footnote{Studies of $Z\to\ell_\alpha\ell_\beta$ with cLFV considering non-degenerate HNL mass spectra exist, as in Ref.~\cite{PilaftsisN}. However, this analysis is performed within a different mass-generation framework.. Only upon completing our manuscript did we become aware of Ref.~\cite{Teixeira}, where the authors consider a couple of non-degenerate HNL-mass ``textures.'' Our approach differs from theirs in that we consider a much broader range of possible non-degenerate HNL mass spectra.}. Our results indicate that the allowed regions for the branching ratios of cLFV decays $Z\to\ell_\alpha\ell_\beta$ change in the presence of HNL mass degeneration, but the restrictions established by non-unitarity constraints and the preservation  of perturbativity avoid larger contributions to the branching ratios, which then turn out to be comparable to those obtained in the case of mass degeneration.}
\\

The rest of the paper has been organized as follows: in Section~\ref{calculation}, we execute the calculation of the whole set of Feynman diagrams contributing to the decay $Z\to\ell_\alpha\ell_\beta$, in the framework of the ISSM; then, in Section~\ref{numbers}, our numerical estimations are presented, analyzed, and discussed, in the light of current and future experimental sensitivity; our paper is concluded with a summary, presented in Section~\ref{summary}.

%%%%%%%%%%%%%%%
%%%%%%%%%%%%%%%
%%%%%%%%%%%%%%%
%%%%%%%%%%%%%%%
%%%%%%%%%%%%%%%

\section{Analytic calculation of $Z\to\ell_\alpha\ell_\beta$}
\label{calculation}
The present section contains the details of the analytic calculation of the one-loop contributions to the cLFV decays $Z\to\ell_\alpha\ell_\beta$, generated by virtual Majorana neutral leptons with masses emerged from the ISSM, which has been briefly discussed in Section~\ref{intro}. However, such previous discussion is general and fails to bear all those aspects necessary to follow the calculation. Therefore, we start this section by discussing the missing material. Then, we fully address the main calculation, after which a discussion of analytic results is carried out. 

%%%%%%%%%%%%%%%
%%%%%%%%%%%%%%%
%%%%%%%%%%%%%%%
%%%%%%%%%%%%%%%
%%%%%%%%%%%%%%%

\subsection{Further theoretical aspects}
\label{moretheory}
The occurrence of the renormalizable terms shown in Eq.~\eqref{renmassterms} determines the couplings of the mass-eigenspinor neutral leptons, both light ones and heavy ones, with the SM particle content. Throughout this subsection, those couplings relevant for the phenomenological calculation are given. We start by establishing the notation required for our ulterior discussion. As explained before, there are nine mass-eigenspinor fields associated to neutral leptons, three of which correspond to the light neutrinos $n_j$ and the rest associated to the HNLs $N_j$ and $X_j$. From now on, whenever we speak of HNLs, and their related quantities, the following notation is used:
\begin{equation}
f_j
=
\left\{
\begin{array}{lr}
N_1,N_2,N_3, & j=1,2,3,
\vspace{0.2cm}
\\
X_1,X_2,X_3, & j=4,5,6.
\end{array}
\right.
\end{equation}
If we refer to the whole set of neutral leptons, comprehending light ones and heavy ones, we utilize the generic notation $\psi_j$, with
\begin{equation}
\psi_j
=
\left\{
\begin{array}{cc}
n_1,n_2,n_3, & j=1,2,3,
\vspace{0.2cm}
\\
N_1,N_2,N_3, & j=4,5,6,
\vspace{0.2cm}
\\
X_1,X_2,X_3, & j=7,8,9,
\end{array}
\right.
\end{equation}
and define the $9\times1$ matrix $\psi$ by its entries: $( \psi )_j=\psi_j$. We also have the $3\times1$ matrix $\ell=\big( \ell_e\hspace{0.2cm}\ell_\mu\hspace{0.2cm}\ell_\tau \big)^\textrm{T}$, with $\ell_\alpha$ the $\alpha$-flavored charged lepton. 
\\

We write the $9\times9$ block-diagonalization matrix $U$, Eq.~\eqref{omegaUV}, as
\begin{equation}
U
=
\left(
\begin{array}{cc}
U_{11} & U_{12}
\vspace{0.2cm}
\\
U_{21} & U_{22}
\end{array}
\right),
\end{equation}
where $U_{jk}$ are block matrices, of which we have two square matrices $U_{11}$ and $U_{22}$, the former $3\times3$ sized and the latter $6\times6$ sized. Moreover, the size of $U_{12}$ is $3\times6$, whereas $U_{21}$ is $6\times3$ sized. We then use these block matrices and the blocks constituting $V$, Eq.~\eqref{Vmatrix}, to define the matrices 
\begin{equation}
\begin{array}{lccr}
{\cal B}_n=\big(U_{11}U^*_\textrm{PMNS}\big)^*, && {\cal B}_f=\big( U_{12}\tilde{V} \big)^*,
\end{array}
\label{Bsdefs}
\end{equation}
which we put together to write the $3\times9$ matrix
\textcolor{Black}{
\begin{equation}
{\cal B}
=
\big(
\begin{array}{ccc}
{\cal B}_n &,& {\cal B}_f
\end{array}
\big)
=
\big( 
\begin{array}{ccc}
{\bf 1}_3 &,& 0_{3\times6}
\end{array}
\big)\Omega^*,
\label{Bdef}
\end{equation}
where $0_{n\times m}$ denotes the $n\times m$ zero matrix. }The ${\cal B}$ matrix fulfills
\begin{equation}
%\begin{array}{lccr}
{\cal B}{\cal B}^\dag={\bf 1}_3, 
\label{Bproperties}
\end{equation}
%%%%%
\begin{equation}
{\cal B}^\dag{\cal B}={\cal C}=
\left(
\begin{array}{cc}
\mathcal{C}_{nn} & \mathcal{C}_{nf} 
\vspace{0.2cm}
\\
\mathcal{C}_{fn} & \mathcal{C}_{ff}
\end{array}
\right),
\label{Bproperties2}
\end{equation}
where ${\cal C}$ is $9\times9$ sized and Hermitian. Notice that Eq.~\eqref{Bproperties2} displays an expression of ${\cal C}$ in terms of matrix blocks, in which $C_{nn}$ is $3\times3$, $C_{ff}$ is $6\times6$, whereas the sizes of ${\cal C}_{nf}$ and ${\cal C}_{fn}$ are $3\times6$ and $6\times3$, respectively. We point out the following useful expression for the matrix $\mathcal{C}$, in terms of the unitary diagonalization matrix $\Omega$:
\begin{equation}
\mathcal{C}
=\Omega^{\rm T}
\left(
\begin{array}{cc}
{\bf 1}_3 & 0_{3\times6}
\vspace{0.2cm}
\\
0_{6\times3} & 0_{6\times6}
\end{array}
\right)\Omega^*.
\label{Cmamalona}
\end{equation}
Eq.~\eqref{Cmamalona} can be used to straightforwardly prove the further properties
\begin{equation}
\mathcal{C}^2=\mathcal{C},
\end{equation}
%%%%%%%%%%
\begin{equation}
\mathcal{B}\mathcal{C}=\mathcal{B},
\label{BCprop}
\end{equation}
%%%%%%%%%%
\textcolor{Black}{
\begin{equation}
\mathcal{B}M_\psi\mathcal{C}^*M_\psi\mathcal{B}^\dag=0,
\label{BMCMBprop}
\end{equation}
where $M_\psi$, defined in Eq~\eqref{numassmatrix}, is the $9\times9$ mass matrix corresponding to the whole set of neutral leptons, so its entries are $\big( M_\psi \big)_{jk}=\delta_{jk}m_{\psi_k}$.}
\\

With the necessary notation and definitions already provided, we now show the couplings to be taken into account for the calculation of loop contributions. We have the Lagrangians
\begin{equation}
{\cal L}_{Wn\ell}=\frac{-g}{\sqrt{2}}W^-_\mu\overline{\ell}\,\mathcal{B}\,\gamma^\mu P_L\psi+{\rm H.c.},
\label{LWnl}
\end{equation}
%%%%%%%%%%
\begin{equation}
\mathcal{L}_{G_Wn\ell}=\frac{-g}{\sqrt{2}m_W}G_W^-\overline{\ell}\big( M_\ell\,\mathcal{B}\,P_L-\mathcal{B}\,M_\psi P_R \big)\psi+{\rm H.c.},
\label{LGnl}
\end{equation}
%%%%%%%%%%%
\begin{equation}
\mathcal{L}_{Znn}=\frac{-g_Z}{2}Z_\mu\overline{\psi}\gamma^\mu\big( \mathcal{C}P_L-\mathcal{C}^*P_R \big)\psi.
\label{LZnn}
\end{equation}
About Eq.~\eqref{LGnl}, the factor $M_\ell$ is the $3\times3$ diagonalized charged-lepton mass matrix, with entries $\big( M_\ell \big)_{\alpha\beta}=\delta_{\alpha\beta}\,m_{\ell_\beta}$. Moreover, the definition 
\begin{equation}
g_Z=\frac{g}{2\cos\theta_{\rm w}},
\end{equation}
with $\theta_{\rm w}$ the weak mixing angle, has been used to write Eq.~\eqref{LZnn}. 

%%%%%%%%%%%%%%%
%%%%%%%%%%%%%%%
%%%%%%%%%%%%%%%
%%%%%%%%%%%%%%%
%%%%%%%%%%%%%%%

\subsection{Neutral lepton contributions at one loop}
\label{gencalc}
\begin{figure}[ht]
\center
\includegraphics[width=6cm]{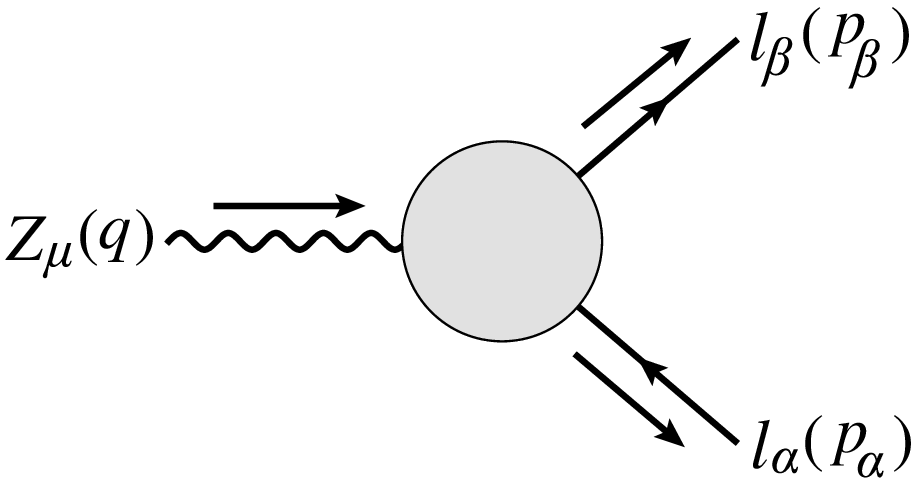},
\caption{\label{Ztoll} Conventions for the $Z\to\ell_\alpha\ell_\beta$ amplitude.}
\end{figure}

Now we calculate the one-loop contributions from virtual light neutrinos $n_j$ and HNLs $N_j$, $X_j$ to the 2-body cLFV $Z$-boson decays. For starters, in the context of the technique of Feynman diagrams~\cite{Feynman}, we follow the conventions provided in Fig.~\ref{Ztoll},
from which the amplitude for the decay $Z\to\ell_\alpha\ell_\beta$ is written as $\mathcal{M}=\overline{u}(p_\beta)\,\Gamma_\mu^{\beta\alpha}\,v(p_\alpha)\epsilon^\mu(q)$, where $u(p_\beta)$ and $v(p_\alpha)$ are momentum-space Dirac spinors, $\epsilon^\mu(q)$ is the polarization 4-vector, and $\Gamma_\mu^{\beta\alpha}$ stands for the vertex function corresponding to the process of interest. Since all the external fermion lines are, by definition, taken on shell, such a vertex function has the Lorentz-covariant parametrization~\cite{HIRSS,NPR,BGS}
\begin{eqnarray}
&&
\Gamma_\mu^{\beta\alpha}=-ig_Z
\Big(
\gamma_\mu\big( f_\textrm{V}^{\beta\alpha}(q^2)+f_\textrm{A}^{\beta\alpha}(q^2)\,\gamma_5 \big)
\nonumber \\
&&
\hspace{1cm}
+i\sigma_{\mu\nu}q^\nu\big( f_\textrm{M}^{\beta\alpha}(q^2)+f_\textrm{E}^{\beta\alpha}\big( q^2 \big)\gamma_5 \big)
\Big),
\label{wempar}
\end{eqnarray}
where $f_\textrm{V}^{\beta\alpha}$ and $f_\textrm{A}^{\beta\alpha}$ parametrize the ``weak-vector current'' and the ``weak-vector-axial current'', respectively, whereas $f_\textrm{M}^{\beta\alpha}$ is the ``weak-magnetic form factor'' and $f_\textrm{E}^{\beta\alpha}$ is known as the ``weak-electric form factor''. The set of all these Lorentz scalars receives the name ``transition weak-electromagnetic form factors''. While Eq.~\eqref{wempar} has its electromagnetic analogue, corresponding to a framework in which the external $Z$ boson (see Fig.~\ref{Ztoll}) is replaced by a photon-field line, note that the $Z$-boson case comes with no restrictions from Ward identities~\cite{Ward}, since the external gauge boson does not abide by electromagnetic gauge invariance. In the case of the electromagnetic analogue of the vertex function $\Gamma_\mu^{\beta\alpha}$, fulfillment of the Ward identity imposes the exact cancellation of both the electromagnetic vector-current from factor and the electromagnetic axial-current form factor, as opposed to what happens in the case at hand. 
\\

Using the couplings given in Eqs.~\eqref{LWnl} to \eqref{LZnn}, the whole set of contributing Feynman diagrams is constructed. While each individual diagram can bear gauge dependence, the sum of all these diagrams must yield gauge-independent contributions to the transition weak-electromagnetic form factors, which define the on-shell amplitude. Therefore, a specific gauge can be picked for the calculation. Moreover, note that, in general, the choice of the gauge determines which diagrams actually exist and thus contribute. For instance, the unitary gauge, defined by the absence of pseudo-Goldstone-boson fields, avoids diagrams in which these fields participate. Also, the gauge-fixing functions characterizing the so-called ``non-linear gauge''~\cite{Shore,EiWu,DDW,MeTo} produce a cancellation of certain couplings among pseudo-Goldstone bosons and gauge fields~\cite{MeTo,NoTo1,GNT,AbTa}, consequently forbidding the occurrence of certain diagrams. For the present work, we have opted for the Feynman-t' Hooft gauge, obtained in the linear gauge-fixing approach~\cite{FLS} by taking the gauge-fixing parameter to be 1. While under this gauge choice the number of contributing diagrams is larger than in the two instances we just cited, it has the advantage of yielding simpler gauge-boson propagators. Once we have the full set of contributing diagrams, we divide the new-physics contributions to $Z\to\ell_\alpha\ell_\beta$ into two parts, which constitute the total contribution to the parametrization given in Eq.~\eqref{wempar}: 
\begin{equation}
\Gamma_\mu^{\beta\alpha}=\sum_{j=1}^9\Gamma_\mu^{\beta\alpha,j}+\sum_{j=1}^9\sum_{k=1}^9\Gamma_\mu^{\beta\alpha,jk},
\label{Gptpluszt}
\end{equation}
with the individual contributions $\Gamma_\mu^{\beta\alpha,j}$ and $\Gamma_\mu^{\beta\alpha,jk}$ diagrammatically expressed as
\begin{eqnarray}
&&
\Gamma_\mu^{\beta\alpha,j}=
\begin{gathered}
\vspace{-0.15cm}
\includegraphics[width=1.6cm]{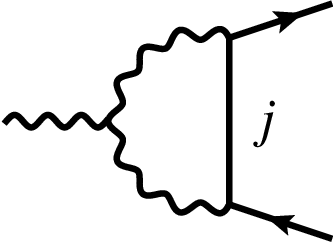}
\end{gathered}
+
\begin{gathered}
\vspace{-0.15cm}
\includegraphics[width=1.6cm]{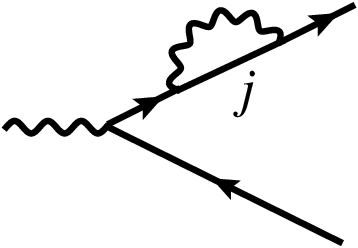}
\end{gathered}
%\nonumber \\ && \hspace{0.78cm}
+
\begin{gathered}
\vspace{-0.15cm}
\includegraphics[width=1.6cm]{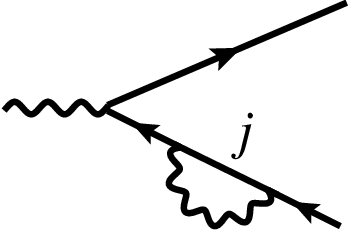}
\end{gathered}
\nonumber \\ \nonumber \\ && \hspace{1cm}
+
\begin{gathered}
\vspace{-0.15cm}
\includegraphics[width=1.6cm]{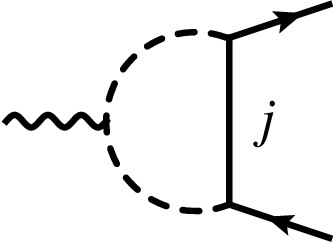}
\end{gathered}
%\nonumber \\ && \hspace{0.78cm}
+
\begin{gathered}
\vspace{-0.15cm}
\includegraphics[width=1.6cm]{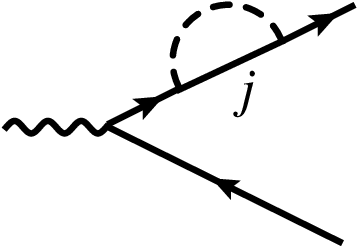}
\end{gathered}
+
\begin{gathered}
\vspace{-0.15cm}
\includegraphics[width=1.6cm]{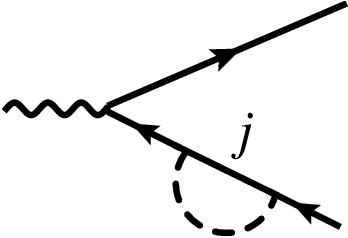}
\end{gathered}
\nonumber \\ \nonumber \\ && \hspace{1cm}
+
\begin{gathered}
\vspace{-0.15cm}
\includegraphics[width=1.6cm]{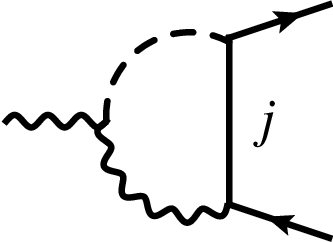}
\end{gathered}
+
\begin{gathered}
\vspace{-0.15cm}
\includegraphics[width=1.6cm]{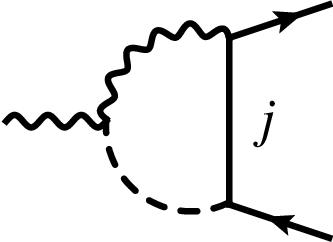}
\end{gathered}
\,,
\label{gdiags}
\end{eqnarray}
%%%%%%%%%%
\begin{equation}
\Gamma_\mu^{\beta\alpha,jk}=
\begin{gathered}
\vspace{-0.15cm}
\includegraphics[width=1.6cm]{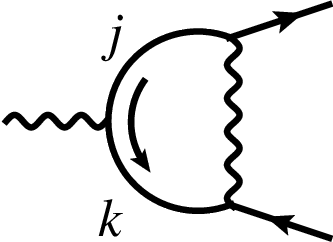}
\end{gathered}
+
\begin{gathered}
\vspace{-0.15cm}
\includegraphics[width=1.6cm]{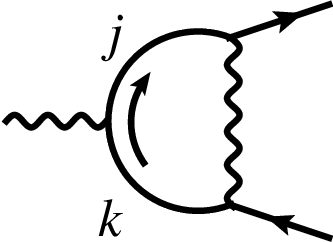}
\end{gathered}
+
\begin{gathered}
\vspace{-0.15cm}
\includegraphics[width=1.6cm]{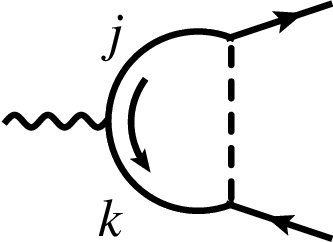}
\end{gathered}
+
\begin{gathered}
\vspace{-0.15cm}
\includegraphics[width=1.6cm]{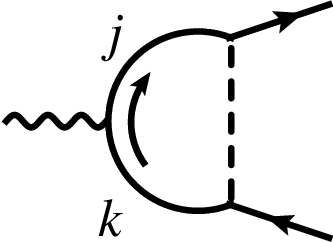}
\end{gathered}
\,.
\label{zdiags}
\end{equation}
The internal lines in all the diagrams of Eqs.~\eqref{gdiags} and \eqref{zdiags} correspond to $W$ bosons (wavy lines), to pseudo-Goldstone bosons $G_W$ (dashed lines), or to neutral leptons $\psi_j=n_j,N_j,X_j$ (solid lines, with no arrows).  The indices $j,k$, in each diagram, briefly denote neutral-lepton fields $\psi_j,\psi_k$. Then, as it can be appreciated from Eqs.~\eqref{gdiags} and \eqref{zdiags}, the sums in Eq.~\eqref{Gptpluszt} run over all the neutral leptons $\psi_j$. An aspect worth of comment, regarding the diagrams that comprise the contribution $\Gamma_\mu^{\beta\alpha,jk}$ displayed in Eq.~\eqref{zdiags}, is related to the Majorana nature of the neutral leptons $\psi_j$. Differences among the Dirac and the Majorana descriptions emerge at different levels, as it is the case of the Feynman rules for Majorana fermions~\cite{GlZr,DEHK}, which allow, in general, for larger sets of diagrams in comparison with the Dirac case. In particular, if we express the $\mathcal{L}_{Z\nu\nu}$ Lagrangian, Eq.~\eqref{LZnn}, as $\mathcal{L}_{Z\nu\nu}=-\sum_{k=1}^9\sum_{j=1}^9iZ_\mu\overline{\psi_k}\,\zeta^\mu_{kj}\psi_j$, besides the Feynman rule
\begin{equation}
\begin{gathered}
\vspace{-0.15cm}
\includegraphics[width=2.5cm]{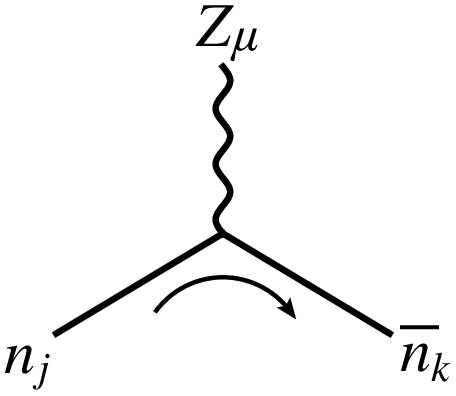}
\end{gathered}
=\zeta^\mu_{kj},
\label{FeynDirac}
\end{equation}
the Feynman rule
\begin{equation}
\begin{gathered}
\vspace{-0.15cm}
\includegraphics[width=2.5cm]{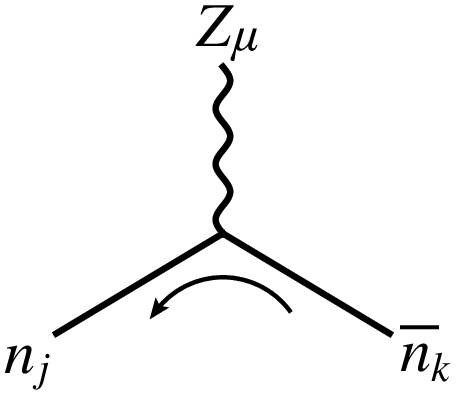}
\end{gathered}
=C\zeta^{\mu\textrm{T}}_{jk}C^{-1},
\label{FeynMajorana}
\end{equation}
exclusive of Majorana fermions, must be taken into account. For the last equation, we have used the symbol $C$ to denote the charge-conjugation matrix. The arrows off neutrino lines in Eqs.~\eqref{FeynDirac}-\eqref{FeynMajorana}, which point out a distinction among these Feynman rules, denote a reference fermion flow aimed at establishing an orientation for fermion chains~\cite{DEHK}. The occurrence of this couple of Feynman rules dictates that for each diagram in which a $Z\psi\psi$
coupling appears, two diagrams, one per each of the Feynman rules displayed in Eqs.~\eqref{FeynDirac}-\eqref{FeynMajorana}, must be set and taken into account. Then notice that both the properties of the charge-conjugation matrix and the structure of the matrix factor $\zeta^\mu_{kj}$ conspire to yield $C\zeta_{jk}^{\mu\textrm{T}}C^{-1}=\zeta^\mu_{kj}$, thus meaning that any two diagrams distinguished only by this vertex are equal to each other, in which case, Eq.~\eqref{zdiags} can be simply written as
\begin{equation}
\Gamma_\mu^{\beta\alpha,jk}=
2\bigg(
\hspace{0.1cm}
\begin{gathered}
\vspace{-0.15cm}
\includegraphics[width=1.6cm]{diag1n}
\end{gathered}
+
\begin{gathered}
\vspace{-0.15cm}
\includegraphics[width=1.6cm]{diag3n}
\end{gathered}
\hspace{0.2cm}
\bigg).
\label{finalMajorana}
\end{equation}
\\

The presence of 4-momentum integrals, associated to loops in diagrams, comes along with latent ultraviolet divergences. Since the SM does not include cLFV tree-level $Z\ell_\alpha\ell_\beta$ couplings, loop-level contributions to such an interaction must be ultraviolet finite, in accordance with renormalization theory, as we are exclusively dealing with renormalizable couplings. Note that, given some physical process, each Feynman loop diagram contributing to it can be, in principle, ultraviolet divergent, but the sum constituting the total contribution, at some loop order, must be free of these divergences, thus meaning that fine cancellations occur when adding together all the contributing diagrams. In order to give a proper treatment to the ultraviolet divergences presumably emerging from the one-loop $Z\ell_\alpha\ell_\beta$ diagrams, the calculation is carried out by using the dimensional-regularization method~\cite{BoGi,tHVe}, in which loop integrals are performed in $D$ spacetime dimensions, with $D\ne4$. The implementation of dimensional regularization involves the replacement $\int\frac{d^4k}{(2\pi)^4}\to\mu_{\rm R}^{4-D}\int\frac{d^Dk}{(2\pi)^D}$, where $\mu_{\rm R}$, known as the ``renormalization scale'', is a quantity with units of mass. At some stage, an analytic continuation is introduced, in which $D$ is assumed to be a complex quantity, in terms of which $\epsilon=4-D$ is defined. In this context, loop integrals are calculated in $\epsilon\to0$. For the calculation of the vertex-function contributions from the Feynman diagrams, we follow the tensor-reduction method, in which the resultant analytic expressions are expressed in terms of Passarino-Veltman scalar functions~\cite{PaVe,DeSt}. While alternative methods to deal with loop contributions, other than tensor reduction, exist, this approach is convenient due to its implementation in software tools. In particular, implementations for the packages \textsc{Feyncalc}~\cite{MBD,SMO1,SMO2,SMO3} and \textsc{Package-X}~\cite{Patel}, used for the present calculation, are available. The $N$-point Passarino-Veltman scalar function, or $N$-point scalar function for short, is defined, in the dimensional-regularization approach, as
\begin{equation}
T^{(N)}_0\big( \textbf{p},\textbf{m} \big)=\frac{\big( 2\pi\mu_\textrm{R} \big)^{4-D}}{i\pi^2}
\int d^Dk\,\prod_{A=0}^{N-1}\big( (k+p_A)^2-m_A^2 \big)^{-1}.
\label{PaVescalar}
\end{equation}
To write down Eq.~\eqref{PaVescalar}, the notation $\textbf{p}=\big( p_1,p_2,\ldots,p_{N-1} \big)$ and $\textbf{m}=\big( m_0,m_1,\ldots, m_{N-1} \big)$ has been used. Moreover, in this general integral, $p_0=0$. An inspection of Eq.~\eqref{PaVescalar}, in the light of its ultraviolet degree of divergence, leads to the conclusion that only 1- and 2-point scalar function are ultraviolet divergent, whereas for $N\geqslant3$ all the $N$-point scalar functions are free of such divergencies. In the case of our calculation, the only source of ultraviolet divergencies are 2-point functions. 
\\

We have found it convenient, for the sake of simplicity, to disregard the masses of charged leptons, appearing in final states, from our calculations. Under these circumstances, we note that the relations $f_{\rm M}^{\beta\alpha}(q^2)=0, \hspace{0.5cm} f_\textrm{E}^{\beta\alpha}(q^2)=0$, and $f_\textrm{A}^{\beta\alpha}(q^2)=-f_\textrm{V}^{\beta\alpha}(q^2)$ hold, so that the vertex function displayed in Eq.~\eqref{wempar} is simply written as
\begin{equation}
\Gamma_\mu^{\beta\alpha}=-ig_Zf_\textrm{A}^{\beta\alpha}(q^2)\gamma_\mu\big( {\bf 1}_4-\gamma_5 \big),
\label{vfunctionapprox}
\end{equation}
then straightforwardly yielding the branching ratio
\begin{equation}
\textrm{Br}\big( Z\to\ell_\alpha\ell_\beta \big)=\frac{g_Z^2m_Z}{6\pi\Gamma_Z^\textrm{tot.}}\big| f_\textrm{A}^{\beta\alpha} \big|^2,
\label{generalBR}
\end{equation}
with $\Gamma_Z^\textrm{tot.}$ denoting the $Z$-boson total decay rate. We express the only remaining form-factor in Eq.~\eqref{vfunctionapprox} as the sum 
\begin{equation}
f_\textrm{A}^{\beta\alpha}=f_{\textrm{A},1}^{\beta\alpha}+f_{\textrm{A},2}^{\beta\alpha}\,,
\label{fAasasum}
\end{equation}
where $f_{\textrm{A},1}^{\beta\alpha}$ is the contribution generated by the diagrams of Eq.~\eqref{gdiags}, in which only one virtual neutral lepton participates, whereas $f_{\textrm{A},2}^{\beta\alpha}$ is the outcome of the diagrams displayed in Eq.~\eqref{finalMajorana}, characterized by two neutral-lepton loop lines. These contributions can be written as
\begin{equation}
f_{\textrm{A},1}^{\beta\alpha}=
%\frac{g^2c_{\rm w}^2}{16\pi^2m_W^4}
\kappa_1
\sum_{j=1}^9\mathcal{B}_{\beta j}\mathcal{B}_{\alpha j}^*f_1(m_{\psi_j}),
\label{fA1}
\end{equation}
%%%%%%%%%%
\begin{eqnarray}
&&
f_{\textrm{A},2}^{\beta\alpha}=
%\frac{g^2}{64\pi^2m_W^2}
\kappa_2
\sum_{k=1}^9\sum_{j=1}^9
\Big(
\mathcal{B}_{\beta k}\mathcal{C}_{kj}\mathcal{B}_{\alpha j}^*f_2(m_{\psi_k},m_{\psi_j})
\nonumber \\ && \hspace{1cm}
+\mathcal{B}_{\beta k}m_k\mathcal{C}_{kj}^*m_j\mathcal{B}_{\alpha j}^*\tilde{f}_2(m_{\psi_k},m_{\psi_j})
\Big),
\label{fA2}
\end{eqnarray}
where $\kappa_1=\frac{g^2_Z}{4\pi^2m_Z^4}$ and $\kappa_2=\frac{g_Z^2}{16\pi^2m_Z^2}$ have been defined. As indicated by the notation, the functions $f_1(m_{\psi_j})$, $f_2(m_{\psi_j},m_{\psi_k})$, and $\tilde{f}_2(m_{\psi_j},m_{\psi_k})$ only depend on the masses of neutral leptons $\psi_j$ (recall we have neglected the masses of charged leptons). Also, due to this, the distinction among the different initial- and final-state charged leptons is solely determined by the matrix entries of $\mathcal{B}$. The dimensional-regularization method allows one to carry out the separations $f_{\textrm{A},1}^{\beta\alpha}=f^{\beta\alpha}_{1,\textrm{fin.}}+f^{\beta\alpha}_{1,\textrm{div.}}$ and $f_{\textrm{A},2}^{\beta\alpha}=f^{\beta\alpha}_{2,\textrm{fin.}}+f^{\beta\alpha}_{2,\textrm{div.}}+\tilde{f}^{\beta\alpha}_{2,\textrm{div.}}$, where $f^{\beta\alpha}_{1,\textrm{fin.}}$ and $f^{\beta\alpha}_{2,\textrm{fin.}}$ are finite, in the ultraviolet sense. Ultraviolet divergences, on the other hand, are comprised by the terms $f^{\beta\alpha}_{1,\textrm{div.}}$, $f^{\beta\alpha}_{2,\textrm{div.}}$, and $\tilde{f}^{\beta\alpha}_{2,\textrm{div.}}$, which are independent of neutral-lepton masses $m_{\psi_j}$. The consideration of this feature allows us to write the divergent parts as
\begin{equation}
f_{1,\textrm{div.}}^{\beta\alpha}\propto \big( {\cal B}\mathcal{B}^\dag \big)_{\beta\alpha}
\Big(
\Delta_\epsilon-\log\frac{\mu_\textrm{R}^2}{m_W^2}
\Big),
\label{f1div}
\end{equation}
%%%%%%%%%%
\begin{equation}
f_{2,\textrm{div.}}^{\beta\alpha}\propto\big( \mathcal{BCB^\dag} \big)_{\beta\alpha}
\Big(
\Delta_\epsilon-\log\frac{\mu_\textrm{R}^2}{m_W^2}
\Big),
\label{f2div}
\end{equation}
%%%%%%%%%%
\begin{equation}
\tilde{f}_{2,\textrm{div.}}^{\beta\alpha}\propto\big( \mathcal{B}M_\psi\mathcal{C}^*M_\psi\mathcal{B}^\dag \big)_{\beta\alpha}
\Big(
\Delta_\epsilon-\log\frac{\mu_\textrm{R}^2}{m_W^2}
\Big),
\label{tildef2div}
\end{equation}
where $\Delta_\epsilon=\frac{2}{\epsilon}-\gamma_{\rm EM}+\log4\pi$, with $\gamma_\textrm{EM}$ the Euler-Mascheroni constant. From Eq.~\eqref{Bproperties}, we note that $\big( \mathcal{B}\mathcal{B}^\dag \big)_{\beta\alpha}=\delta_{\beta\alpha}=0$, since $\beta\ne\alpha$. Furthermore, Eq.~\eqref{BCprop} leads us to $\big( \mathcal{B}\mathcal{C}\mathcal{B}^\dag \big)_{\beta\alpha}=\delta_{\beta\alpha}=0$. Finally, we have the property $\mathcal{B}M_\psi\mathcal{C}^*M_\psi\mathcal{B}^\dag=0$, given above, in Eq.~\eqref{BMCMBprop}. That Eqs.~\eqref{f1div}-\eqref{tildef2div} are satisfied means that the amplitude $\mathcal{M}$, for the decay $Z\to\ell_\alpha\ell_\beta$, is finite, in the ultraviolet sense, and independent of the renormalization scale $\mu_\textrm{R}$. 
\\

\textcolor{Black}{
Let us remark that the afore-discussed cancellation of ultraviolet divergences bears a resemblance to the GIM mechanism. Thinking of this, we can anticipate the further elimination of finite contributions by the same token. From Eq.~\eqref{Bproperties}, we find the relation $\sum_{j=1}^3\mathcal{B}_{\beta X_j}\mathcal{B}^*_{\alpha X_j}=-\sum_{j=1}^3\big( \mathcal{B}_{\beta n_j}\mathcal{B}^*_{\alpha n_j}+\mathcal{B}_{\beta N_j}\mathcal{B}^*_{\alpha N_j} \big)$, which holds since we are assuming that $\beta\ne\alpha$. In what follows, we take $m_{N_j}=m_{X_j}$, for any $j=1,2,3$, in accordance with the near-degeneracy condition among HNL pairs $\big( N_j,X_j \big)$. Let us also comment that the enormous difference among the masses of light neutrinos and the $W$-boson mass, both associated to virtual lines running in the loops of contributing Feynman diagrams, makes, from a quantitative perspective, the role of the former, in the $f_1$, $f_2$, $\tilde{f}_2$ contributing factors, much less important than the significance of the latter, so we can neglect light-neutrino masses and just take $m_{n_j}=0$, which we do from here on. The explicit expressions for the functions $f_1$, $f_2$, and $\tilde{f}_2$, once both light-neutrino and charged-lepton masses have been neglected, can be found in Appendix~\ref{analyticexpressions}. However, note that we are not assuming that light neutrinos are massless; we only have taken advantage of quite-suppressed subdominant contributions from light-neutrino masses to $f_1$, $f_2$, $\tilde{f}_2$. This observation is opportune and relevant, as the ISSM relation for the masses of light neutrinos, shown in Eq~\eqref{issrelation}, still holds. Under the afore-discussed assumptions on the neutral-lepton mass spectrum, we find that $f_{\textrm{A},1}^{\beta\alpha}$, given in Eq.~\eqref{fA1}, acquires the form
\begin{equation}
f_{\textrm{A,1}}^{\beta\alpha}=\kappa_1\sum_{j=1}^3\mathcal{B}_{\beta n_j}\mathcal{B}^*_{\alpha n_j}\big( f_1(0)-f_1(m_{f_j}) \big).
\label{f1indeltasv2}
\end{equation}
Moreover, we note that Eqs.~\eqref{BCprop} and \eqref{BMCMBprop} allow one to write $f_{\textrm{A},2}^{\beta\alpha}$, Eq.~\eqref{fA2}, as
\begin{eqnarray}
&&
f_{\textrm{A},2}^{\beta\alpha}=\kappa_2
\Big(
\sum_{k=1}^3\sum_{j=1}^3\mathcal{B}_{\beta n_k}\mathcal{C}_{n_kn_j}\mathcal{B}^*_{\alpha n_j}\Delta f_2^{kj}(0,0)
\nonumber \\ && \hspace{1cm}
+\sum_{k=1}^3\sum_{j=1}^6\mathcal{B}_{\beta n_k}\mathcal{C}_{n_kf_j}\mathcal{B}^*_{\alpha f_j}\Delta f_2^{kj}(0,m_{f_j})
\nonumber \\ && \hspace{1cm}
+\sum_{k=1}^6\sum_{j=1}^3\mathcal{B}_{\beta f_k}\mathcal{C}_{f_k n_j}\mathcal{B}^*_{\alpha n_j}\Delta f_2^{kj}(m_{f_k},0)
\Big),
\label{f2indeltasv2}
\end{eqnarray}
for which the difference $\Delta f_2^{kj}(m_{\psi_i},m_{\psi_r})=f_2(m_{\psi_i},m_{\psi_r})-f(m_{f_k},m_{f_j})$ has been defined. We also note that usage of $m_{n_j}=0$ and $m_{N_j}=m_{X_j}$ has come along with a complete cancellation of the second term in Eq.~\eqref{fA2}, in which the factors $\tilde{f}_2(m_{\psi_k},m_{\psi_j})$ are involved. The differences $f_1(0)-f_1(m_{f_j})$, appearing in Eq.~\eqref{f1indeltasv2}, and $\Delta f_2^{kj}(m_{\psi_i},m_{\psi_{r}})$, featured by Eq.~\eqref{f2indeltasv2}, implement cancellations among terms, in particular eliminating those independent of neutral-lepton masses, which include the ultraviolet-divergent contributions comprising $f_{1,\textrm{div.}}^{\beta\alpha}$ and $f_{2,\textrm{div.}}^{\beta\alpha}$. In this way, consistently with our discussion of the last paragraph, Eqs.~\eqref{f1indeltasv2} and \eqref{f2indeltasv2} turn out to be free of ultraviolet divergences. Furthermore, taking $m_{N_1}=\Lambda$, then reasonably assuming that $m_{N_2}\propto\Lambda$ and $m_{N_3}\propto\Lambda$, and keeping in mind that $m_{N_j}=m_{X_j}$, we have been able to verify that the limits
\begin{equation}
\lim_{\Lambda\to\infty} f_{\textrm{A}}^{\beta\alpha}=0
\hspace{0.3cm}
\Rightarrow
\hspace{0.3cm}
\lim_{\Lambda\to\infty} \textrm{Br}\big( Z\to\ell_\alpha\ell_\beta \big)=0,
\end{equation}
are fulfilled, which is consistent with the decoupling of new-physics contributions as the high-energy scale $\Lambda$ grows, in accordance with the decoupling theorem~\cite{ApCa}.} 
\\

%%%%%%%%%%%%%%%
%%%%%%%%%%%%%%%
%%%%%%%%%%%%%%%
%%%%%%%%%%%%%%%
%%%%%%%%%%%%%%%

\subsection{Neutral-lepton couplings and non-unitarity effects}
\label{etadiscussion}
In the simplest neutrino-mass-generating schemes, such as the SM extended by Yukawa terms involving three right-handed singlet neutrino fields or the SM effective field theory~\cite{LLR,BuWy,Wudka,BaLe} endowed with the Weinberg operator~\cite{Weinbergoperator}, the charged-current terms are expressed, in the basis of mass eigenspinor neutrino fields, as
\begin{equation}
\mathcal{L}_{Wn\ell}^\textrm{SM}=\frac{-g}{\sqrt{2}}W^-_\mu\overline{\ell}\,\gamma_\mu P_L U_\textrm{PMNS}\,n+\textrm{H.c.}
\label{ccSM}
\end{equation}
The transformation $\nu=U_\textrm{PMNS}\,n$, given by the PMNS mixing matrix then defines the neutrino flavor basis, characterized by the flavor-neutrino fields $(\nu)_\alpha=\nu_e,\nu_\mu,\nu_\tau$. On the other hand, in more intricate neutrino mass mechanisms in which HNLs are involved, as the one followed for the present investigation, the charged currents exclusively involving light neutrinos are written, in the neutrino-mass basis, as
\begin{equation}
\mathcal{L}_{Wn\ell}^\textrm{light}=\frac{-g}{\sqrt{2}}W^-_\mu\overline{\ell}\,\gamma^\mu P_L\mathcal{B}_nn+\textrm{H.c.},
\end{equation}
which directly follows from Eq.~\eqref{LWnl}. From its definition, the matrix $\mathcal{B}$ is not restricted to be unitary, so the light-neutrino flavor basis is no more defined by a unitary transformation, a consequence of the presence of further neutral leptons. In the context of the ISSM, in which the $3\times3$ matrix
\begin{equation}
\xi=m_\textrm{D}M^{-1}
\label{xidef}
\end{equation}
is very small, 
the charged currents are expressed as
\begin{equation}
\mathcal{L}^\textrm{light}_{Wn\ell}=\frac{-g}{\sqrt{2}}W^-_\mu\overline{\ell}\,\gamma^\mu P_L\big( \textbf{1}_3-\eta \big)U_\textrm{PMNS}\,n+\textrm{H.c.},
\label{ccISS}
\end{equation}
where we have defined the Hermitian $3\times3$ matrix
\begin{equation}
\eta=\frac{1}{2}\xi\xi^\dag,
\label{etadef}
\end{equation}
whose entries are $\eta_{\beta\alpha}$. Comparison of Eqs.~\eqref{ccSM} and \eqref{ccISS} lead us to conclude that $\eta$ characterizes non-unitarity effects of light-neutrino mixing, which, if detected, would point towards the presence of HNLs associated to physics beyond the SM. No compelling evidence of such non-unitarity effects have been ever measured, though studies focused on them, such as those of Refs~\cite{GoNo,FHL,BFHLMN,FGT}, have established upper constraints on the entries of $\eta$ that are as restrictive as $10^{-6}$. 
\\

\textcolor{Black}{
We have discussed above how the couplings of the neutral leptons $n_j$, $N_j$, and $X_j$ are determined by the matrices $\mathcal{B}$ and $\mathcal{C}$, defined in Eqs.~\eqref{Bsdefs} and \eqref{Bproperties2}, respectively. Already from Eqs.~\eqref{fA1} and \eqref{fA2}, we note that the contributions determining $f_\textrm{A}^{\beta\alpha}$, and thus the $Z\to\ell_\alpha\ell_\beta$ amplitude $\Gamma_\mu^{\beta\alpha}$, displayed in Eq.~\eqref{vfunctionapprox}, are given in terms of the matrices $\mathcal{B}$ and $\mathcal{C}$. In order to connect our analysis of contributions with non-unitarity effects, we note that both the $\mathcal{B}$ and $\mathcal{C}$ matrices can be expressed in terms of the non-unitarity matrix $\eta$, defined in Eq.~\eqref{etadef}. We use Eqs.~\eqref{Bsdefs} and \eqref{Bproperties2}, together with Eqs.~\eqref{omegaUV}-\eqref{Vmatrix} and Eq.~\eqref{etadef} to express $\mathcal{B}$ and $\mathcal{C}$ as 
\begin{equation}
\mathcal{B}_n=\big( \textbf{1}_3-\eta+\frac{3}{2}\eta^2 \big)U_\textrm{PMNS},
\label{Bninetas}
\end{equation}
%%%%%%%%%%
\begin{equation}
\mathcal{B}_f=\frac{1}{\sqrt{2}}\big( \textbf{1}_3-\eta \big)\xi
\left(
\begin{array}{ccc}
i\cdot\textbf{1}_3 &,& \textbf{1}_3
\end{array}
\right),
\end{equation}
%%%%%%%%%%
\begin{equation}
\mathcal{C}_{nn}=U_\textrm{PMNS}^\dag\big( \textbf{1}_3-2\eta+4\eta^2 \big)U_\textrm{PMNS},
\end{equation}
%%%%%%%%%%
\begin{equation}
\mathcal{C}_{nf}=\frac{1}{\sqrt{2}}U_\textrm{PMNS}^\dag\big( \textbf{1}_3-2\eta \big)\xi
\left(
\begin{array}{ccc}
i\cdot\textbf{1}_3 &,& \textbf{1}_3
\end{array}
\right),
\end{equation}
%%%%%%%%%%
\begin{equation}
\mathcal{C}_{fn}=\frac{1}{\sqrt{2}}
\left(
\begin{array}{c}
-i\cdot\textbf{1}_3
\vspace{0.2cm}
\\
\textbf{1}_3
\end{array}
\right)\xi^\dag\big( i\cdot\textbf{1}_3-2\eta \big)U_\textrm{PMNS},
\end{equation}
%%%%%%%%%%
\begin{equation}
\mathcal{C}_{ff}=\frac{1}{2}
\left(
\begin{array}{c}
-i\cdot\textbf{1}_3
\vspace{0.2cm}
\\
\textbf{1}_3
\end{array}
\right)
\xi^\dag\big( \textbf{1}_3-2\eta \big)\xi
\left(
\begin{array}{ccc}
i\cdot\textbf{1}_3 &,& \textbf{1}_3
\end{array}
\right),
\label{Cffinetas}
\end{equation}
at the second order in the non-unitarity matrix $\eta$.} We have previously commented that the mass of any heavy neutrino $N_j$ is almost the same as that of the corresponding HNL $X_j$, that is, $m_{N_j}\approx m_{X_j}$ for any $j=1,2,3$. In fact, such a difference amounts to the small factor $\mu_{R}+\mu_{S}$, determined by the energy scale $v_\sigma$. However, in general, the heavy-neutrino mass spectrum $\big\{m_{N_1}, m_{N_2}, m_{N_3}\big\}$, and thus the set $\big\{ m_{X_1},m_{X_2},m_{X_3} \big\}$, is, in principle, not restricted to be degenerate or even near-degenerate. Below, when performing quantitative estimations, we consider two cases for the set of masses of HNLs: (1) the set of masses is degenerate, so $m_{f_j}=m_{f_k}$ for any $j,k=1,2,3,4,5,6$; and (2) this set is general, not constrained to be degenerate.\\

\subsection{Degenerate HNL masses}
\label{degenerate}
In this subsection, we consider a context in which the masses of the heavy neutrinos comprise a degenerate set, so $m_{N_j}=m_{N_k}$ for any pair $j,k=1,2,3$. Then notice that $m_{N_j}\approx m_{X_j}$, whichever $j=1,2,3$ is taken. Taking both aspects into account, we consider, in practice, that the six HNLs have the same mass, namely, $m_{f_j}=m_N$, for any $j=1,2,3,4,5,6$ and where $m_N$ is some reference mass, used to characterize the heavy masses. \textcolor{Black}{Following the assumption of HNL mass degeneracy, along with Eqs.~\eqref{Bninetas} to \eqref{Cffinetas}, we find that the contributions $f_{\textrm{A},1}^{\beta\alpha}$ and $f_{\textrm{A},2}^{\beta\alpha}$, displayed in Eqs.~\eqref{f1indeltasv2} and \eqref{f2indeltasv2}, are given in terms of the non-unitarity matrix $\eta$, by
\begin{equation}
f_{\textrm{A},1}^{\beta\alpha}=-2\kappa_1\big( \eta-2\eta^2 \big)_{\beta\alpha}\big( f_1(0)-f_1(m_N) \big),
\label{fA1indegenerate}
\end{equation}
%%%%%%%%%%
\begin{eqnarray}
&&
f_{\textrm{A},2}^{\beta\alpha}=-2\kappa_2\Big(
\eta_{\beta\alpha}\big( \Delta f_2(0,0,m_N)+\Delta f_2(0,m_N,0) \big)
\nonumber \\ && \hspace{1cm}
-(\eta^2)_{\beta\alpha}\big( 3\Delta f_2(0,0,m_N)+3\Delta f_2(0,m_N,0)
\nonumber \\ && \hspace{0.8cm}
+\Delta f_2(m_N,0,m_N)
+\Delta f_2(m_N,m_N,0) \big)
\Big),
\label{fA2indegenerate}
\end{eqnarray}
with the difference $\Delta f_2(m_1,m_2,m_3)=f_2(m_1,m_1)-f_2(m_2,m_3)$ defined.
}
\\

\section{Estimations and analyses}
\label{numbers}
The occurrence of cLFV is forbidden in the framework of the SM, in which neutrinos are considered massless. However, the mixing of massive neutrinos, involved in neutrino oscillations, allows for non-preservation of charged-lepton flavor, thus implying that processes such as $\ell_\alpha\to\ell_\beta\gamma$ and $\ell_\alpha\to\ell_\beta\ell_\omega\ell_\omega$, absent in the SM phenomenology, can actually take place and might indeed show up in forthcoming experimental studies. About this, note that a constraint as stringent as $\textrm{Br}\big( \mu\to e\gamma \big)_\textrm{MEG II}<3.1\times10^{-13}$ has already been established from the MEG II update~\cite{MEG2}. The Belle and BaBar Collaborations have searched for cLFV tau-lepton decays $\tau\to e\gamma$ and $\tau\to\mu\gamma$, establishing upper bounds of order $10^{-8}$ on the corresponding branching ratios~\cite{BaBar,Belle}. Long ago, the SINDRUM Collaboration performed an investigation in search for the cLFV decay $\mu\to3e$, then finding the restriction $\textrm{Br}\big( \mu\to3e \big)_\textrm{SINDRUM}<1.0\times10^{-12}$~\cite{SINDRUM}. The participation of unknown new physics, beyond the SM, in the non-conservation of charged-lepton flavor is expected, so the exploration of this phenomenon at the theoretical, phenomenological, and experimental levels is of utmost importance. The cLFV decays $Z\to\ell_\alpha\ell_\beta$, which are the focus of the present work, have never been measured, while experimental searches and analyses have established upper bounds on their branching ratios. In the minimally extended SM, the GIM mechanism dramatically suppresses contributions, then yielding branching ratios as tiny as $10^{-60}$~\cite{IJR}. A recent investigation, by the ATLAS Collaboration, has used data from proton-proton collisions at a center-of-mass energy of $13\,\textrm{TeV}$ to search for cLFV through the process $Z\to\mu e$, then arriving at the conclusion that $\textrm{Br}\big( Z\to e\mu \big)_\textrm{ATLAS}<2.62\times10^{-7}$, at the $95\%$ confidence level~\cite{ATLASZtoemu}. Ref.~\cite{ATLASZtotaualgo}, also by the ATLAS Collaboration, has addressed the cLFV decays $Z\to e\tau$ and $Z\to\mu\tau$, performing a search from proton-proton collision data obtained at a center-of-mass energy of $13\,\textrm{TeV}$. From the joint consideration of their analysis and a previous study by the same experimental group~\cite{ATLASZcLFVdecays}, they reach the constraints $\textrm{Br}\big( Z\to\tau e \big)_\textrm{ATLAS}<5.0\times10^{-6}$ and $\textrm{Br}\big( Z\to\tau\mu \big)_\textrm{ATLAS}<6.5\times10^{-6}$, both given at the $95\%$ confidence level. While, as shown above, experimental constraints on $Z\to\ell_\alpha\ell_\beta$ are currently available, a step forward has been taken in studies in which the expected sensitivity of future facilities to these processes has been estimated. This is, for instance, the case of Ref.~\cite{FCCeebounds}, where a sensitivity of the FCC-ee to $Z\to\mu e$ at the $10^{-10}$ level is claimed to be achievable. The author of Ref.~\cite{FCCeebounds} also points out that a sensitivity at the $10^{-9}$ level of the FCC-ee to both $Z\to\tau e$ and $Z\to\tau\mu$ could be reached. Similar expected sensitivities to $Z\to\ell_\alpha\ell_\beta$ have been estimated for the Circular Electron Positron Collider (CEPC)~\cite{CEPCbounds}. We summarize this discussion in Table~\ref{Ztollboundsdata}.
\begin{table}[ht]
\center
\begin{tabular}{c|c|c|c}
Process & ATLAS (current) & FCC-ee & CEPC
\\\hline\hline
$Z\to e\mu$ & $<2.62\times10^{-7}$ & $10^{-10}-10^{-8}$ & $10^{-9}$
\\
$Z\to e\tau$ & $<5.0\times10^{-6}$ & $10^{-9}$ & -
\\
$Z\to\mu\tau$ & $<6.5\times10^{-6}$ & $10^{-9}$ & $10^{-9}$
\end{tabular}
\caption{\label{Ztollboundsdata} cLFV $Z$ decays. Current bounds~\cite{ATLASZtoemu,ATLASZtotaualgo}; FCC-ee expected sensitivity~\cite{FCCeebounds}; and CEPC expected sensitivity~\cite{CEPCbounds}.}
\end{table}
\\

In view of the confirmation of neutrino mass and mixing, the determination of the values of the light-neutrino masses has become a main objective of experimental searches. The role played by squared-mass differences $\Delta m_{jk}^2=m_{n_j}^2-m_{n_k}^2$ in neutrino oscillations has been used by a number of experimental collaborations to determine the values~\cite{PDG,SKDeltam12,ORCADeltam32,DBDeltam32,SKDeltam32,ICDeltam32,T2KDeltam32,DBoDeltam32otro,NOvADeltam32,MINOSDeltam32,RENODeltam32}
\begin{equation}
\textrm{NH}
\left\{
\begin{array}{l}
\Delta m_{21}^2=\big( 7.53\pm0.18 \big)\times10^{-5}\,\textrm{eV}^2,
\vspace{0.2cm}\\
\Delta m_{32}^2=\big( 2.455\pm0.028 \big)\times10^{-3}\,\textrm{eV}^2,
\end{array}
\right.
\label{NHexp}
\end{equation}
%%%%%%%%%%
\begin{equation}
\textrm{IH}
\left\{
\begin{array}{l}
\Delta m_{21}^2=\big( 7.53\pm0.18 \big)\times10^{-5}\,\textrm{eV}^2,
\vspace{0.2cm}\\
\Delta m_{32}^2=\big( -2.529\pm0.029 \big)\times10^{-3}\,\textrm{eV}^2,
\end{array}
\right.
\label{IHexp}
\end{equation}
delivered by the Particle Data Group~\cite{PDG}. Here, NH and IH are the acronyms for ``normal hierarchy'' and ``inverted hierarchy', respectively. And, by the way, the signs for $\Delta m_{32}^2$ in Eqs.~\eqref{NHexp} and \eqref{IHexp} differ because whether the mass of the neutrino $n_3$ is the largest one, corresponding to normal hierarchy, or the smallest one, if the hierarchy is inverted, remains to be determined. Settling which neutrino-mass hierarchy is the one actually occurring in nature is one of the goals of the physics programs of future experimental facilities, such as the Deep Underground Neutrino Experiment~\cite{DUNEDeltam32}, the DUNE, and the Hyper-Kamiokande~\cite{HKDeltam32}. On the other hand, the Jiangmen Underground Neutrino Observatory, better known as JUNO, has just started taking data, with the purpose of solving this issue, and their first results are already available~\cite{JUNODeltam32}. Moreover, the absolute neutrino mass scale also remains unknown. Searches for the elusive neutrinoless double beta decay, allowed only if neutrinos are of Majorana type, have been taken profit of to upper bound the effective Majorana neutrino mass, 
defined as $m^\textrm{eff}_{\nu_e}=\big| \sum_j\big( U_\textrm{PMNS} \big)^2_{ej}\,m_{n_j} \big|$, from which restrictions at the sub-eV level have been set, at the $90\%$ confidence level~\cite{CUORE,GERDA,KamLANDZen}. Cosmological data have been used to put the upper limit $\sum_j m_{n_j}<0.072\,\textrm{eV}$~\cite{PlanckCollab,DESICollab}, at the $95\%$ confidence level, on the sum of neutrino masses, though notice that this constraint relies on cosmological assumptions. The KATRIN Collaboration has recently reported an upper bound on light-neutrino masses, which is independent of both cosmological  models and whether light neutrinos are of Dirac or Majorana type. This group claims, in Ref.~\cite{KATRIN}, that the restriction $m_n<0.45\,\textrm{eV}$, at the $90\%$ confidence level, is fulfilled. 
%\textcolor{Red}{From here on, our quantitative analysis and discussion follow the KATRIN result. And talking about our calculation, let us point out that no appreciable differences among the results in the NH and those in the IH have been observed. Therefore, our numerical evaluations are presented under the assumption of normally ordered masses of light neutrinos. Given the squared-mass differences $\Delta m_{31}^2=\Delta m_{32}^2+\Delta m_{21}^2$ and $\Delta m_{32}^2$, the masses $m_{n_1}$ and $m_{n_2}$ can be written down, in the NH case, as
%\begin{equation}
%m_{n_1}=\sqrt{m_{n_3}^2-\Delta m_{31}^2},
%\label{mn1numerical}
%\end{equation}
%%%%%%%%%%
%\begin{equation}
%m_{n_2}=\sqrt{m_{n_3}^2-\Delta m_{32}^2}.
%\label{mn2numerical}
%\end{equation}
%Taking the KATRIN upper bound as reference, we consider values of the $m_{n_3}$ mass ranging within
%\begin{equation}
%\sqrt{\Delta m_{31}^2}\leqslant m_{n_3}\leqslant %0.45\,\textrm{eV}.
%\label{mn3numerical}
%\end{equation}
%}
\\

The PMNS neutrino mixing matrix, $U_\textrm{PMNS}$, characterizing neutrino mixing in minimal extensions of the neutrino sector of the SM, is determined by 4 or 6 parameters, depending on whether these neutrinos correspond to Dirac or Majorana fields. If neutrinos are Majorana fermions, as it is the case of the ISSM, considered for the present investigation, this matrix can be expressed as $U_\textrm{PMNS}=U_\textrm{D}U_\textrm{M}$, where
\begin{widetext}
\begin{equation}
U_\textrm{D}=
\left(
\begin{array}{ccc}
c_{12}c_{13} & s_{12}s_{13} & e^{-i\delta_\textrm{D}}s_{13}
 \vspace{0.2cm} \\
-s_{12}c_{23}-e^{i\delta_\textrm{D}}c_{12}s_{23}s_{13} & c_{12}c_{23}-e^{i\delta\textrm{D}}s_{12}s_{23}s_{13} & s_{23}c_{13}
 \vspace{0.2cm} \\
s_{12}s_{23}-e^{i\delta_\textrm{D}}c_{12}c_{23}s_{13} & -c_{12}s_{23}-e^{i\delta_\textrm{D}}s_{12}c_{23}s_{13} & c_{23}c_{13}
\end{array}
\right),
\end{equation}
\end{widetext}
for which the short-hand notation $s_{jk}=\sin\theta_{jk}$ and $c_{jk}=\cos\theta_{jk}$ has been used to refer to sines and cosines of mixing angles $\theta_{12}$, $\theta_{23}$, and $\theta_{13}$. The values for the mixing angles, as recommended by the Particle Data Group~\cite{PDG}, are given by the relations
\begin{eqnarray}
\sin^2\theta_{12}=0.307\pm0.013,
\\
\sin^2\theta_{23}=0.546\pm0.0021,
\\
\sin^2\theta_{13}=0.0220\pm0.0007.
\end{eqnarray}
In order to establish these values, the Particle Data Group has considered the experimental results reported in Refs.~\cite{SKDeltam12,ORCADeltam32,SKDeltam32,ICDeltam32,T2KDeltam32,NOvADeltam32,MINOSDeltam32,DBDeltam32,NOvAmixing,DoubleChoozmixing,RENOmixing1,RENODeltam32}. The ``Dirac phase'', another parameter of the matrix $U_\textrm{D}$, has been addressed by Refs.~\cite{PDG,SKDeltam32,T2KDeltam32,NOvADeltam32}, in which the conclusion that
\begin{equation}
\delta_\textrm{D}=1.23\pm0.21\pi\,\textrm{rad}
\end{equation}
has been reached. If neutrinos end up being of Dirac type, the parameters of $U_\textrm{D}$ are just enough to characterize light-neutrino mixing. Nonetheless, in a world in which light neutrinos are Majorana fermions, the mixing would require two further parameters, namely, a couple of ``Majorana phases'', $\phi_1$ and $\phi_2$, which are part of the matrix
\begin{equation}
U_\textrm{M}=
\left(
\begin{array}{ccc}
1 & 0 & 0
 \vspace{0.2cm} \\
0 & e^{i\phi_1} & 0
 \vspace{0.2cm} \\
0 & 0 & e^{i\phi_2}
\end{array}
\right).
\end{equation}
In pursuit of a smaller set of free parameters, our quantitative estimations are carried out by assuming that $\phi_1=0$ and $\phi_2=0$. 
\\

Defined above, in Eq.~\eqref{etadef}, the matrix $\eta$, of non-unitarity effects in the light-neutrino sector, is $3\times3$ sized and Hermitian, thus involving 6 independent entries. In Ref.~\cite{GoNo}, by the authors of the present investigation, the cLFV decays $\ell_\alpha\to\ell_\beta\gamma$ were explored in the framework of the ISSM. That work includes, in particular, a discussion in which, in accordance with the hierarchy condition among energy scales given in Eq.~\eqref{hierarchy}, the limit as $\frac{v_\sigma^2}{v^2}\to0$ (thus $\frac{m_{n_j}^2}{m_W^2}\to0$) is considered and the ratio $\frac{\Lambda^2}{v^2}$ (or $\frac{m_{f_j}^2}{m_W^2}$) is assumed to be very large. Under such circumstances, the branching ratio for the decay process $\ell_\alpha\to\ell_\beta\gamma$ fulfills an appealing relation with the non-unitarity matrix-entry $\eta_{\beta\alpha}$, namely, $\textrm{Br}\big( \ell_\alpha\to\ell_\beta\gamma \big)\propto \big| \eta_{\beta\alpha} \big|^2$. In such a framework, the authors of Ref.~\cite{GoNo} evoked the upper bound $4.2\times10^{-13}$ on the branching ratio for $ \mu\to e\gamma$, reported by the MEG Collaboration in Ref.~\cite{previousMEGbound}. This experimental restriction, pertinent at the time, was then used to set the constraint $\big| \eta_{\mu e} \big|_\textrm{MEG}\lesssim1.14\times10^{-5}$, though the remark was made that the MEG II update would eventually be able to impose the more strict upper limit $6\times10^{-14}$ on this branching ratio~\cite{MEG2future}, then improving the upper bound on $\big| \eta_{\mu e} \big|$ by a factor $\sim\frac{1}{3}$:
\begin{equation}
\big| \eta_{\mu e} \big|_\textrm{future}\lesssim 4.29\times10^{-6}.
\label{etaconstraintfuture}
\end{equation}
It turns out that the MEG II is already operating and has improved their restriction on the $\mu\to e\gamma$ branching ratio, claiming that $\textrm{Br}\big( \mu\to e\gamma \big)_\textrm{MEG II}<3.1\times10^{-13}$~\cite{MEG2}. In view of this event, we also update the result of Ref.~\cite{GoNo} on the non-unitarity matrix-entry $\eta_{\mu e}$:
\begin{equation}
\big| \eta_{\mu e} \big|_\textrm{current}\lesssim9.75\times10^{-6}.
\label{etaconstraint}
\end{equation}
Following the same approach, restrictions on $\eta_{\tau e}$ and $\eta_{\tau\mu}$ can also be derived. 
%However, the results of Ref.~\cite{GoNo} show that $\mu\to e\gamma$ yields the most stringent constraints, as the ISSM contributions to its branching ratio are the ones which might lie closer to experimental sensitivity right now or in the near future. 
%\textcolor{Red}{The later comment is well-timed, since the methodology driving the quantitative analyses performed in the present investigation follows what has been done in Ref.~\cite{GoNo}.}
\\

\textcolor{Black}{As far as the constraints on the rest of non-unitarity parameters are concerned, we follow Ref.~\cite{BFHLMN}, where a comprehensive analysis, aimed at limiting the full $\eta$ matrix has been performed, taking into account a number of electroweak precision observables: determination of the $W$-boson mass; determination of the effective weak mixing angle; five observables measured at the $Z$ pole by the Large Electron Positron Collider, and a CMS determination of the $Z$-boson invisible decay width; five decay ratios that constrain lepton-flavor universality; ten weak decays which bound unitarity of the Cabibbo-Kobayashi-Maskawa mixing matrix; and cLFV observables. Be, nonetheless, aware that the bounds established in Ref.~\cite{BFHLMN} have been set in the framework of the type-1 seesaw mechanism, or, more precisely, in scenarios of this mechanism, which distinguish from each other by the number of HNLs. Even so, we use these constraints as a reference, though note that in the case of the parameter $\big| \eta_{\mu e} \big|$ we rather use the bound derived in Ref.~\cite{GoNo}, which is given in Eq.~\eqref{etaconstraint}.} \textcolor{Black}{The set of limits on $\eta$ used as a benchmark in the present work are summarized by 
\begin{equation}
|\eta|\leqslant\eta_\textrm{max},
\end{equation}
where
\begin{equation}
\eta_\textrm{max}=
\left(
\begin{array}{ccc}
1.3\times10^{-3} & 9.75\times10^{-6} & 8.0\times10^{-4}
\vspace{0.2cm}
\\
9.75\times10^{-6} & 1.1\times10^{-5} & 1.8\times10^{-5}
\vspace{0.2cm}
\\
8.0\times10^{-4} & 1.8\times10^{-5} & 8.1\times10^{-4}
\end{array}
\right).
\label{etamax}
\end{equation}
Let us comment that different sets of values for these constraints hold depending on whether the neutrino-mass ordering is normal or inverted. For each $\eta_\textrm{max}$ entry in Eq.~\eqref{etamax}, we have considered the most stringent limit among those given by the two neutrino mass orderings. 
}
\\

\textcolor{Black}{With the objective of reducing the number of free parameters involved in the quantitative estimations of $\textrm{Br}\big( Z\to\ell_\alpha\ell_\beta \big)$, we recall Eqs.~\eqref{MNmatrix} and \eqref{MXmatrix}, according to which the diagonal HNL mass matrices are given as $M_N\approx M_X\approx M$, with $M$ the matrix that mixes $\nu_R$ and $S_L$ as shown in Eq.~\eqref{renmassterms}. We then assume $M$ to be linked to the high-energy scale $\Lambda$ as
\begin{equation}
M=\Lambda\,\zeta_M.
\label{MinLambda}
\end{equation}
In this context, we also assume that the matrix $\zeta_M$ is diagonal and real, with $( \zeta_M )_{jj}\sim\mathcal{O}(1)$. It is worth remarking that the particular texture $\zeta_M=\textbf{1}_3$ corresponds to a context in which all the HNLs share the same mass: $m_{f_j}=m_N$ for $j=1,2,3,3,4,5,6$.} \textcolor{Black}{We now make the assumption that $m_\textrm{D}=v\hat{Y}\times10^n$ and $\mu_S=v_\sigma\zeta_\mu$, where $\hat{Y}$ and $\zeta_\mu$ are, for now, assumed to be complex matrices, with $| \hat{Y}_{jk} |\sim\mathcal{O}(1)$ and $| ( \zeta_\mu )_{jk} |\sim\mathcal{O}(1)$, whereas $n$ is some natural number which determines the order of magnitude of the Yukawa coupling $m_\textrm{D}$. Keeping all these assumptions in mind, the consideration of HNL-mass degeneration by pairs ($m_{N_j}=m_{X_j}$, for $j=1,2,3$) allows us to write down the ISSM relation for the masses of light neutrinos, Eq.~\eqref{issrelation}, as
\begin{equation}
M_n=
\bigg(
\frac{v_\sigma v^2}{\Lambda^2}\times10^{2n}
\bigg)
U_\textrm{PMNS}^*\hat{Y}\zeta_M^{-1}\zeta_\mu\zeta_M^{-1}\hat{Y}^\textrm{T}U_\textrm{PMNS}^\dag.
\label{ISSMrelationestimations}
\end{equation}
Since, in accordance with the KATRIN bound~\cite{KATRIN}, the masses of light neutrinos are upper-bounded as $M_n<0.45\textrm{eV}$, we can sensibly assume that
\begin{equation}
\frac{v_\sigma v^2}{\Lambda^2}\times10^{2n}\sim1\,\textrm{eV},
\end{equation}
which allows us to quantify the lepton-number-violation scale $v_\sigma$ in terms of how large the energy scales $v$ and $\Lambda$ are. In this context, we note that the non-unitarity matrix $\eta$ gets the form
\begin{equation}
\eta=
\bigg(
\frac{v^2}{2\Lambda^2}\times10^{2n}
\bigg)
\hat{Y}\zeta_M^{-2}\hat{Y}^\textrm{T}.
\label{etainY}
\end{equation}
} \textcolor{Black}{A convenient approach to deal with the $\hat{Y}$ matrix, which is part of Eq.~\eqref{ISSMrelationestimations}, has been suggested in Ref.~\cite{AHMW}. In accordance with that work, we consider a framework in which $\tilde{Y}$ is a real matrix. We then express $\hat{Y}$ as
\begin{equation}
\hat{Y}=
\left(
\begin{array}{c}
\vec{\eta}_e
\vspace{0.2cm}
\\
\vec{\eta}_\mu
\vspace{0.2cm}
\\
\vec{\eta}_\tau
\end{array}
\right)
R,
\label{geometricY}
\end{equation}
where $\vec{\eta}_\alpha$, with $\alpha=e,\mu,\tau$, are 3-dimensional vectors, accommodated as $1\times3$ matrices, whereas $R$ is an orthogonal matrix representing a rotation implemented on these vectors. Eq.~\eqref{geometricY} shall be useful below.}\\

\subsection{Numerical estimations: degenerate HNL masses}
\textcolor{Black}{A discussion on numerical estimations of the branching ratios $\textrm{Br}\big( Z\to\ell_\alpha\ell_\beta \big)$ is presented throughout this subsection, in which the ISSM contributions are considered in a context in which the set of HNL masses is taken to be degenerate. As earlier established, the mass shared in this framework by all the HNLs is denoted by $m_N$. Recall that this degeneracy-mass assumption fits the $\zeta_M$ texture $\zeta_M=\textbf{1}_3$. Taking $m_N=\Lambda$ and using Eq.~\eqref{geometricY}, we write $\eta$ as
\begin{equation}
\eta=
%\bigg(
\frac{v^2}{2m_N^2}\times10^{2n}
%\bigg)
\left(
\begin{array}{ccccc}
|\vec{\eta}_e|^2 && \vec{\eta}_e\cdot\vec{\eta}_\mu && \vec{\eta}_e\cdot\vec{\eta}_\tau
\vspace{0.4cm}
\\
\vec{\eta}_\mu\cdot\vec{\eta}_e && |\vec{\eta}_\mu|^2 && \vec{\eta}_\mu\cdot\vec{\eta}_\tau
\vspace{0.4cm}
\\
\vec{\eta}_\tau\cdot\vec{\eta}_e && \vec{\eta}_\tau\cdot\vec{\eta}_\mu && |\vec{\eta}_\tau|^2
\end{array}
\label{etageometric}
\right).
\end{equation}
In this equation, the parameters determining the non-unitarity effects are the three moduli of the vectors $\vec{\eta}_\alpha$ and the relative angles among them. Moreover, according to Eq.~\eqref{etageometric}, the entries of the non-unitarity matrix are larger or smaller, with respect to each other, depending on how parallel or perpendicular the vectors $\vec{\eta}_\alpha$ are. A further relevant feature of Eq.~\eqref{etageometric} is its independence of the rotation $R$, which holds only in the case of degenerate HNL masses. In fact, this $R$-rotation independence is inherited by the branching ratio $\textrm{Br}\big( Z\to\ell_\alpha\ell_\beta \big)$. To grasp this asseveration, recall that, as shown in Eqs.~\eqref{generalBR} and \eqref{fAasasum}, this branching ratio is determined by the form factor $f_\textrm{A}^{\beta\alpha}=f_{\textrm{A},1}^{\beta\alpha}+f_{\textrm{A},2}^{\beta\alpha}$. In the HNL-mass degenerate scenario, the contributing terms $f_{\textrm{A},1}^{\beta\alpha}$ and $f_{\textrm{A},2}^{\beta\alpha}$, given by Eqs~\eqref{fA1indegenerate} and \eqref{fA2indegenerate}, are solely determined by the non-unitarity matrix $\eta$ and by the HNL mass $m_N$. Since the rotation matrix $R$ has vanished from $\eta$, we arrive at the conclusion that $\textrm{Br}\big( Z\to\ell_\alpha\ell_\beta \big)$ is also $R$ independent as long as a mass-degenerate HNL mass spectrum is assumed.
}
\\

\textcolor{Black}{In this framework of HNL-mass degeneration, we have set $\eta=\eta_\textrm{max}$, with $\eta_\textrm{max}$ given in Eq.~\eqref{etamax}, and then we have plotted the graphs displayed in Fig.~\ref{degenerateetamax},
\begin{figure}[ht]
\centering
\includegraphics[width=8cm]{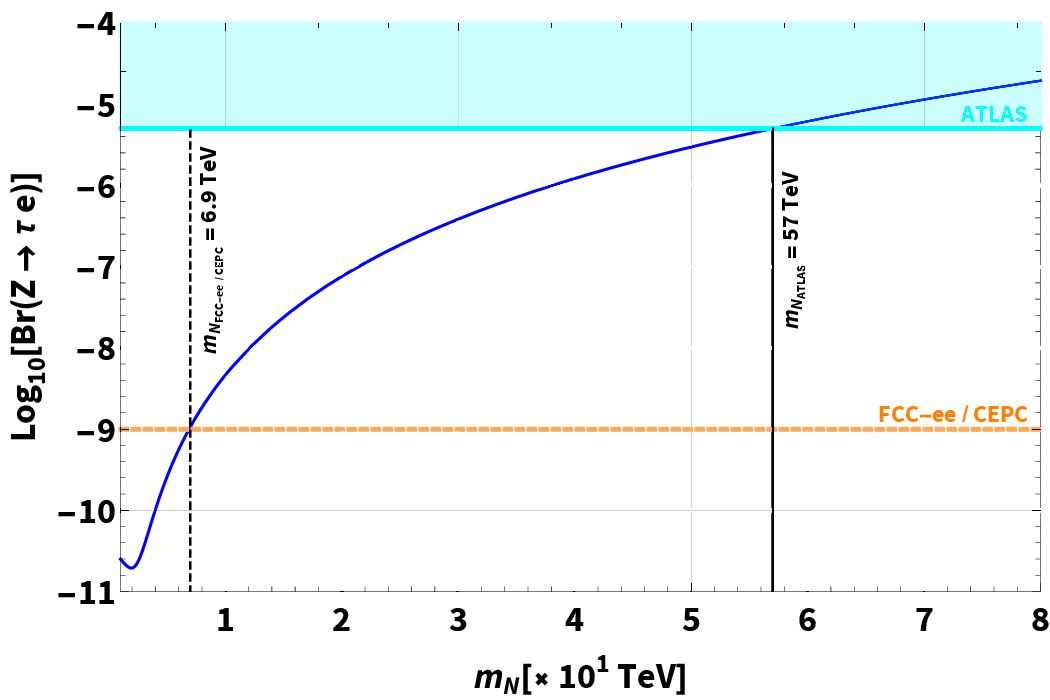}
\vspace{0.2cm}
\\
\includegraphics[width=8cm]{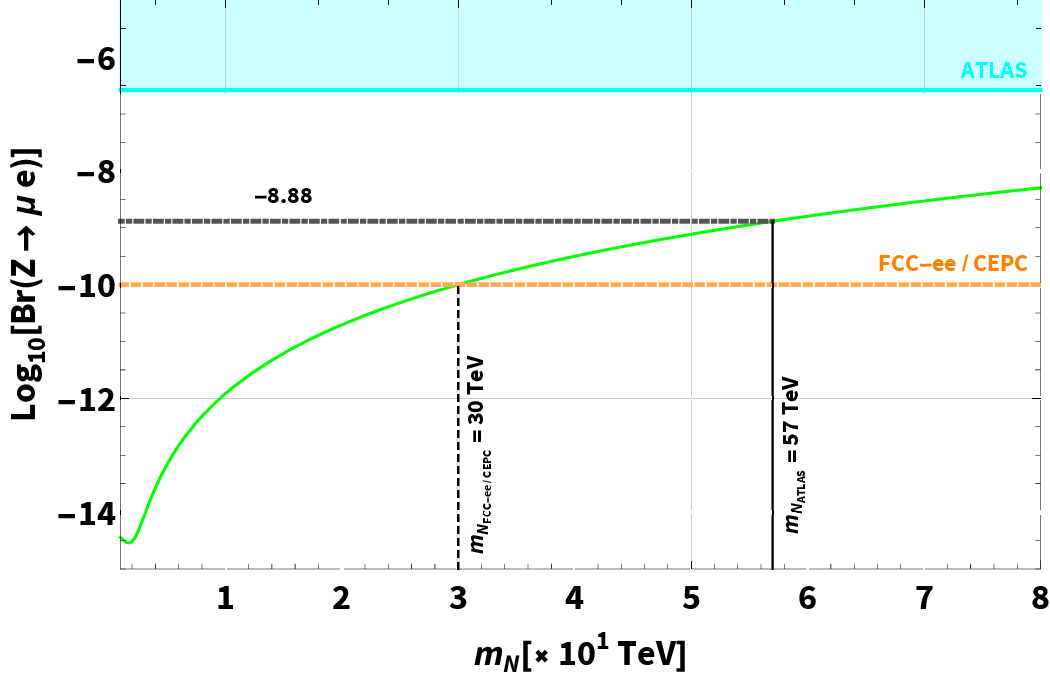}
\vspace{0.2cm}
\\
\includegraphics[width=8cm]{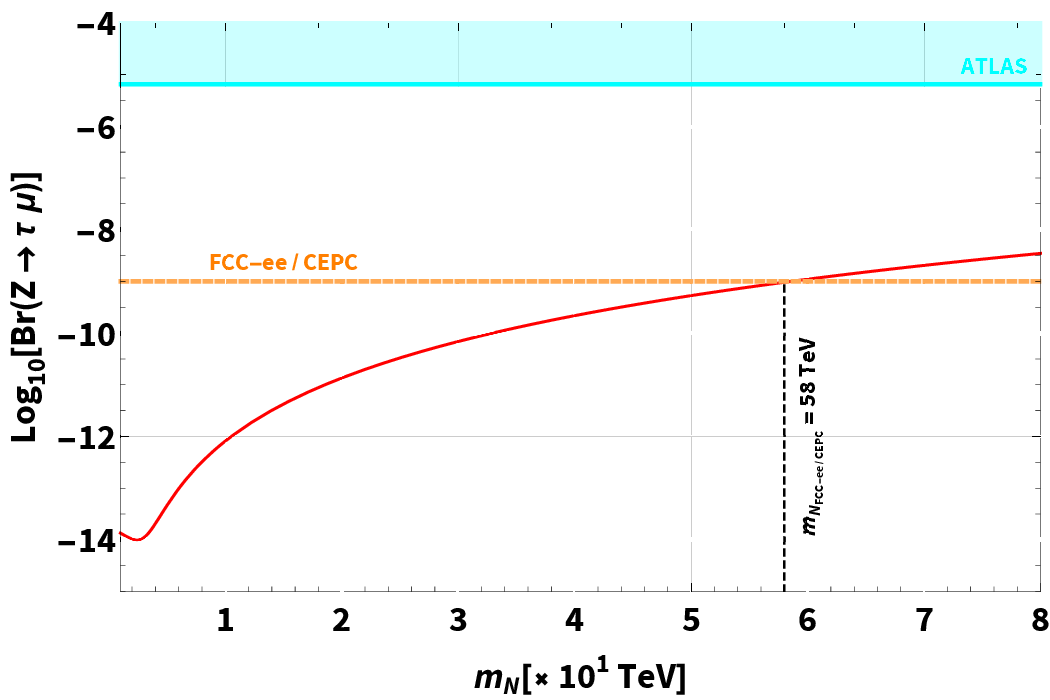}
\caption{\label{degenerateetamax} \textcolor{Black}{Graphs for the cLFV branching ratios $\textrm{Br}\big( Z\to\ell_\alpha\ell_\beta \big)$, in base-10 logarithmic scale, as functions of the degenerate HNL mass $m_N$, with the assumption that $\eta=\eta_\textrm{max}$. The horizontal lines in the top of the graphs represent current experimental sensitivity, whereas dashed horizontal lines stand for projected sensitivity of the FCC-ee and the CEPC to these $Z$-boson decays.}}
\end{figure}
which depict ISSM contributions to branching ratios for $Z$-boson cLFV decays: $\textrm{Br}\big( Z\to\tau e \big)$, shown in the upper panel; $\textrm{Br}\big( Z\to\mu e \big)$ displayed in middle panel; and $\textrm{Br}\big( Z\to\tau\mu \big)$, appearing in the lower panel. All these graphs for the branching-ratio contributions, as functions of the HNL mass $m_N$, have been effectuated in base-10 logarithmic scale, with $1\,\textrm{TeV}\leqslant m_N\leqslant80\,\textrm{TeV}$. 
%While the shape of the plots for the contributions look, at first glance, similar, note that the scale used for $m_N$, running over the horizontal axis and given in TeVs, is particular for each decay. 
All the graphs include a shaded region in their upper regions, aimed at representing those branching-ratio values already excluded by experimental data~\cite{ATLASZtoemu,ATLASZtotaualgo}. Also, a dashed horizontal line has been plotted in each graph in order to show the sensitivity improvement expected for each decay from the FCC-ee and the CEPC~\cite{FCCeebounds,CEPCbounds}. These graphs indicate that $\textrm{Br}\big( Z\to\tau e \big)$ is the most promising one among the branching ratios for the 2-body cLFV $Z$  decay processes. The corresponding graph, shown in the upper panel, includes a vertical solid line, meant to indicate what the $m_N$ value yielding the largest $\textrm{Br}\big( Z\to\tau e \big)$ associated to $\eta=\eta_\textrm{max}$, as established by current experimental sensitivity, is, with this value amounting to $m_N^\textrm{curr.}=57\,\textrm{TeV}$. Analogously, the vertical dashed line represents such an HNL-mass value, but given in accordance with projected future sensitivity. We find that FCC-ee and CEPC will be able to upper bound the HNL mass, through $\textrm{Br}\big( Z\to\tau e \big)$, by $m_N^\textrm{fut.}=6.9\,\textrm{TeV}$. Note, however, that these limits established on $m_N$ correspond to the scenario in which $\eta=\eta_\textrm{max}$. If, on the other hand, $\eta<\eta_\textrm{max}$ values are rather considered, the bounds on $m_N$ become larger. Being $Z\to\tau e$ the process establishing the most stringent restrictions, the constraint $m_N^\textrm{curr.}$, derived on the HNL mass $m_N$ from current experimental data on $\textrm{Br}\big( Z\to\tau e \big)$, can then be used to set more restrictive upper bounds on the further two cLFV branching ratios $\textrm{Br}\big( Z\to\mu e \big)$ and $\textrm{Br}\big( Z\to\tau\mu \big)$. In this manner, we determine the upper limit $\textrm{Br}\big( Z\to\mu e \big)<1.32\times10^{-9}$, represented in the middle panel of Fig.~\ref{degenerateetamax} by the dot-dashed horizontal line. In the case of $Z\to\tau\mu$, the value $m_N^\textrm{curr.}$ sets an upper bound on the corresponding branching ratio which practically coincides with the expected upper limit from FCC-e and CEPC.}\\

\textcolor{Black}{Now we perform a broader quantitative analysis, with the objective of exploring further parameter scenarios beyond the maximal non-unitarity violation characterized by $\eta_\textrm{max}$, Eq.~\eqref{etamax}. To this aim, we use the parametrization
\begin{equation}
\vec{\eta}_e=|\vec{\eta}_e|\,\big( \sqrt{1-x^2},0,x \big),
\label{vecetae}
\end{equation}
%%%%%%%%%%
\begin{equation}
\vec{\eta}_\mu=|\vec{\eta}_\mu|\,\big(0, \sqrt{1-x^2},x \big),
\label{vecetamu}
\end{equation}
%%%%%%%%%%
\begin{equation}
\vec{\eta}_\tau=|\vec{\eta}_\tau|\,\big(\sqrt{1-x^2},x,0 \big),
\label{vecetatau}
\end{equation}
where, for the moment, $0<x<1$. The cases $x=0$ and $x=1$ have been excluded in order to keep $\hat{Y}$ invertible (see Ref.~\cite{AHMW}).} \textcolor{Black}{ Therefore, $\eta$ can be written down as
\begin{widetext}
\begin{equation}
\eta=\frac{v^2}{2m_N^2}\times10^{2n}
\left(
\begin{array}{ccc}
|\vec{\eta}_e|^2 & |\vec{\eta}_e||\vec{\eta}_\mu|x^2 & |\vec{\eta}_e||\vec{\eta}_\tau|(1-x^2)
\vspace{0.3cm} \\
|\vec{\eta}_e||\vec{\eta}_\mu|x^2 & |\vec{\eta}_\mu|^2 & |\vec{\eta}_\mu||\vec{\eta}_\tau|x\sqrt{1-x^2}
\vspace{0.3cm} \\
|\vec{\eta}_e||\vec{\eta}_\tau|(1-x^2) & |\vec{\eta}_\mu||\vec{\eta}_\tau|x\sqrt{1-x^2} & |\vec{\eta}_\tau|^2
\end{array}
\right).
\label{etageometricpar}
\end{equation}
\end{widetext}
Note that, from a geometric viewpoint, the directions of the vectors $\vec{\eta}_\alpha$ and $\vec{\eta}_\beta$ define angles among them, $\theta_{\alpha\beta}$, determined by
\begin{equation}
\cos\theta_{e\mu}=x^2,
\label{thetaemu}
\end{equation} 
%%%%%%%%%%
\begin{equation}
\cos\theta_{\mu\tau}=x\sqrt{1-x^2},
\label{thetamutau}
\end{equation}
%%%%%%%%%%
\begin{equation}
\cos\theta_{\tau e}=1-x^2.
\label{thetataue}
\end{equation}
So, the parameter $x$ is meant to tune the degree of perpendicularness or parallelness of pairs of vectors $\vec{\eta}_\alpha$, and, in this way, control the off-diagonal entries of the non-unitarity matrix $\eta$, in accordance with the expression given in Eq.~\eqref{etageometricpar}. For instance, in $x\to0$ the vectors $\vec{\eta}_e$ and $\vec{\eta}_\tau$ are nearly parallel, whereas $\vec{\eta}_e$ and $\vec{\eta}_\tau$ are both almost perpendicular to $\vec{\eta}_\mu$, so in this scenario the entries $\eta_{e\mu}$ and $\eta_{\mu\tau}$ get suppressed whereas $\eta_{\tau e}$ is enhanced. If, on the other hand, a context in which $x\to1$ is considered, the vectors $\vec{\eta}_e$ and $\vec{\eta}_\mu$ become parallel, while both of them are approximately perpendicular to $\vec{\eta}_\tau$, which comes along with a suppression of the entries $\eta_{\tau e}$, $\eta_{\mu\tau}$ and a larger $\eta_{e\mu}$. Comparison of Eqs.~\eqref{etamax} and \eqref{etageometricpar} then suggests that current experimental data on non-unitarity effects favor a scenario in which $x$ lies closer to $0$. For the numerical estimations, we have looked for a reduction of the number of parameters involved in the expression for the branching ratio, which is in part determined by the non-unitarity matrix $\eta$. We have considered the possibility of decreasing the number of parameters associated to $\eta$ by taking $\eta_\textrm{max}$ as our reference to establish a reference $m_N$ value determined by current constraints on non-unitarity effects. With such an objective, we have defined the HNL reference mass $m_{N_0}$, abiding by the condition $\eta(m_N)\Big|_{m_N=m_{N_0}}=\eta_\textrm{max}$. Then, we find that $\eta$ can be expressed as
%\begin{widetext}
%\begin{equation}
%\eta=\frac{v^2}{2m_N^2}\,|\vec{\eta}_\tau|^2\times10^{2n}
%\left(
%\begin{array}{ccc}
%Primera fila
%\frac{(\eta_\textrm{max})^2_{\tau e}}{(\eta_\textrm{max})_{\tau\tau}^2}
%%%%%
%&
%%%%%
%\frac{(\eta_\textrm{max})_{\tau e}}{(\eta_\textrm{max})_{\tau\tau}}\sqrt{\frac{(\eta_\textrm{max})_{\mu\mu}}{(\eta_\textrm{max})_{\tau\tau}}}x^2
%%%%%
%&
%%%%%
%\frac{(\eta_\textrm{max})_{\tau e}}{(\eta_\textrm{max})_{\tau\tau}}(1-x^2)
%\vspace{0.3cm} \\
%Segunda fila
%\frac{(\eta_\textrm{max})_{\tau e}}{(\eta_\textrm{max})_{\tau\tau}}\sqrt{\frac{(\eta_\textrm{max})_{\mu\mu}}{(\eta_\textrm{max})_{\tau\tau}}}x^2
%%%%%
%&
%%%%%
%\frac{(\eta_\textrm{max})_{\mu\mu}}{(\eta_\textrm{max})_{\tau\tau}}
%%%%%
%&
%%%%%
%\sqrt{\frac{(\eta_\textrm{max})_{\mu\mu}}{(\eta_\textrm{max})_{\tau\tau}}}x\sqrt{1-x^2}
%\vspace{0.3cm} \\
%Tercera fila
%\frac{(\eta_\textrm{max})_{\tau e}}{(\eta_\textrm{max})_{\tau\tau}}(1-x^2)
%%%%%
%&
%%%%%
%\sqrt{\frac{(\eta_\textrm{max})_{\mu\mu}}{(\eta_\textrm{max})_{\tau\tau}}}x\sqrt{1-x^2}
%%%%%
%&
%%%%%
%1
%\end{array}
%\right),
%\label{etainetamax}
%\end{equation}
%\end{widetext}
\begin{widetext}
\begin{equation}
\eta=\frac{v^2}{2m_N^2}\,|\vec{\eta}_\tau|^2\times10^{2n}
\left(
\begin{array}{ccccc}
%Primera fila
r_{\tau e}^2
%%%%%
&&
%%%%%
r_{\tau e}\sqrt{r_{\mu\mu}}\,x^2
%%%%%
&&
%%%%%
r_{\tau e}(1-x^2)
\vspace{0.3cm} \\
%Segunda fila
r_{\tau e}\sqrt{r_{\mu\mu}}\,x^2
%%%%%
&&
%%%%%
r_{\mu\mu}
%%%%%
&&
%%%%%
\sqrt{r_{\mu\mu}}\,x\sqrt{1-x^2}
\vspace{0.3cm} \\
%Tercera fila
r_{\tau e}(1-x^2)
%%%%%
&&
%%%%%
\sqrt{r_{\mu\mu}}\,x\sqrt{1-x^2}
%%%%%
&&
%%%%%
1
\end{array}
\right),
\label{etainetamax}
\end{equation}
\end{widetext}
for which the ratios
\begin{equation}
r_{\alpha\beta}=\frac{(\eta_\textrm{max})_{\alpha\beta}}{(\eta_\textrm{max})_{\tau\tau}}
\end{equation}
have been defined.
%The HNL mass is constrained as $m_{N_0}\leqslant m_N$, with $m_{N_0}$ such that\footnote{\textcolor{RoyalBlue}{As discussed in Appendix~\ref{etaderivationdegenerate}, in practice $\eta(m_N)\big|_{m_N=m_{N_0}}\approx\eta_\textrm{max}$. at $x=0.1944$.}} $\eta(m_N)\big|_{m_N=m_{N_0}}=\eta_\textrm{max}$. 
Also, the tuning parameter $x$ turns out to be restricted as 
\begin{equation}
0<x\leqslant0.1944.
\label{xtuninterval}
\end{equation}
The details of the derivation of this equation can be found in Appendix~\ref{etaderivationdegenerate}. Regarding the condition $\eta(m_N)\Big|_{m_N=m_{N_0}}=\eta_\textrm{max}$, we find it worth commenting that, as discussed in Appendix~\ref{etaderivationdegenerate}, in practice what is really found is that Eq.~\eqref{etainetamax} rather fulfills $\eta(m_N)\Big|_{m_N=m_{N_0}}\approx\eta_\textrm{max}$ at $x=0.1944$. The $\eta$ matrix, as given in Eq.~\eqref{etainetamax}, has three free parameters, namely, the HNL mass $m_N$, the tuning parameter $x$, and the coefficient $n$ setting the order of magnitude of the $m_\textrm{D}$ Yukawa matrix.} \textcolor{Black}{As part of the quantitative analyses, we perform a scan over allowed values for such parameters, taking into account the restriction
\begin{equation}
|\eta|\leqslant (\eta_\textrm{max}).
\end{equation}
Also we take the fixed value $x=0.1944$, for the tuning parameter, in accordance with Eq.~\eqref{xtuninterval}. In the case of $n$, we consider the values $n=-1,0$, taking into account the restriction set by the perturbative regime: 
\begin{equation}
\big|Y^\nu\big|<\sqrt{4\pi},
\label{pertubativitylimit}
\end{equation}
for which $Y^\nu=\hat{Y}\times10^{n}$ has been defined, so that
\begin{equation}
1\leqslant|\vec{\eta}_\tau|\sqrt{1-x^2}\leqslant10,\textrm{ if } n=-1, \Rightarrow Y^\nu\sim10^{-1},
\label{nandeta1}
\end{equation}
%%%%%%%%%%
\begin{equation}
1\leqslant|\vec{\eta}_\tau|\sqrt{1-x^2}\leqslant\sqrt{4\pi},\textrm{ if } n=0,\Rightarrow Y^\nu\sim\mathcal{O}(1).
\label{nandeta2}
\end{equation}
}

\textcolor{Black}{With the previous discussion in mind, we refer the reader to the graphs of Fig.~\ref{broaderanalysis},
\begin{figure}[ht]
\center
\includegraphics[width=7cm]{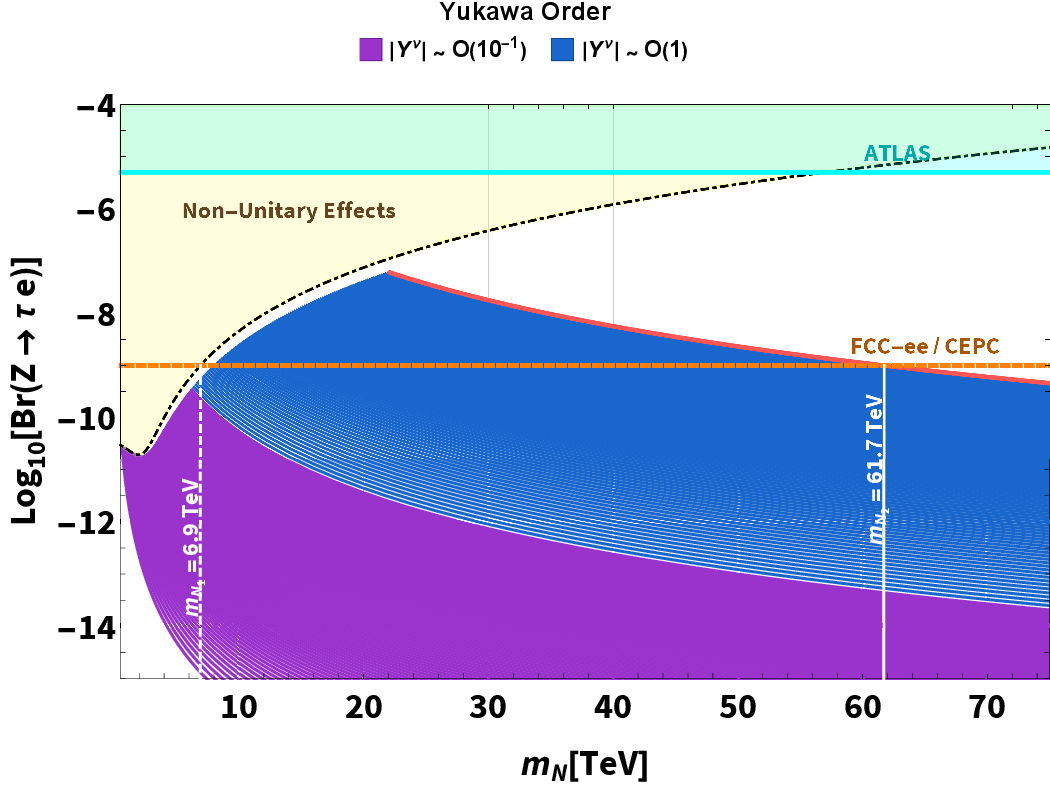}
\vspace{0.5cm}
\\
\includegraphics[width=7cm]{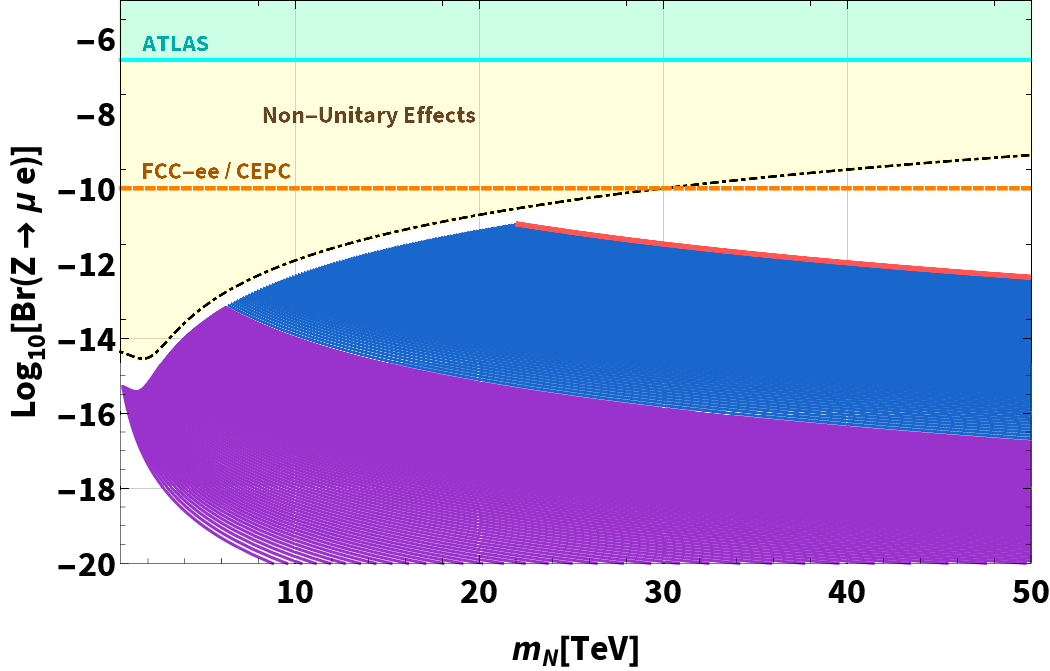}
\vspace{0.5cm}
\\
\includegraphics[width=7cm]{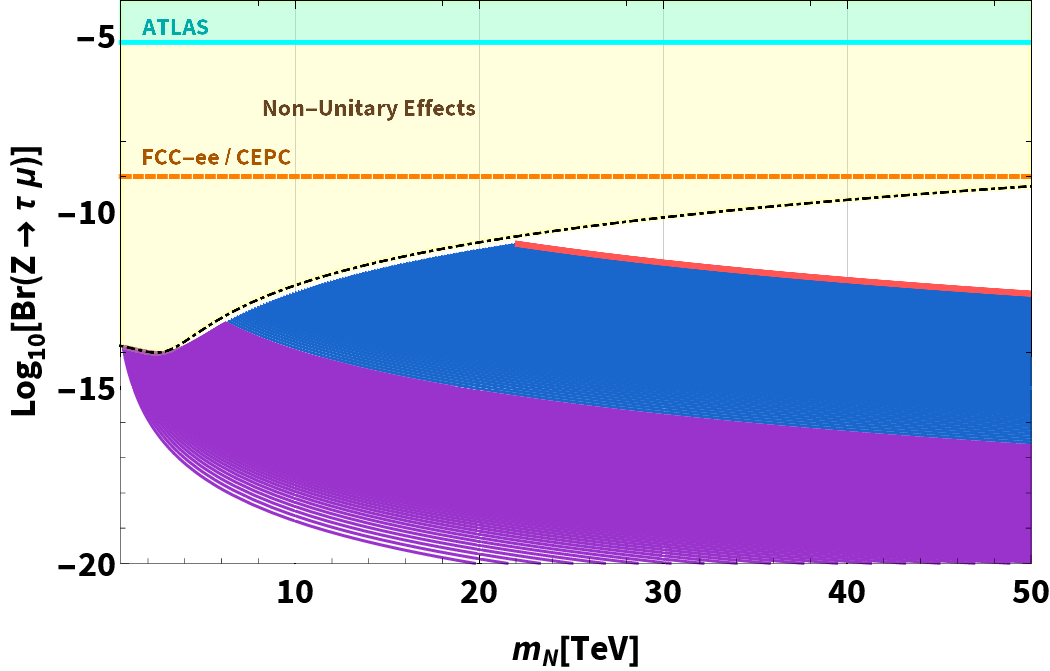}
\caption{\label{broaderanalysis} \textcolor{Black}{Allowed regions for the cLFV 2-body decays $Z\to\ell_\alpha\ell_\beta$ in the scenario of HNL mass degeneration, displayed in the $\big( m_N,\textrm{Br}( Z\to\ell_\alpha\ell_\beta ) \big)$ parameter space. The blue regions correspond to $Y^\nu\sim\mathcal{O}(1)$, whereas the purple regions are associated to $Y^\nu\sim\mathcal{O}(10^{-1})$. These allowed regions are upper-bounded by the validity of the perturbative regime (red solid curves) and by constraints on non-unitarity effects (dot-dashed curves).}}
\end{figure}
which display allowed regions in the parameter space $\big( m_N,\textrm{Br}( Z\to\ell_\alpha\ell_\beta ) \big)$, thus providing a representation of the branching ratios for the cLFV decays $Z\to\ell_\alpha\ell_\beta$. The values along the branching-ratio axes are displayed in base-10 logarithmic scale, whereas the values constituting the $m_N$ axes are within $1\,\textrm{TeV}\leqslant m_N\leqslant75\,\textrm{TeV}$ in the upper panel, where $\textrm{Br}\big( Z\to\tau e \big)$ is shown, and $1\,\textrm{TeV}\leqslant m_N\leqslant50\,\textrm{TeV}$ in the cases of the middle and the lower panels, corresponding to $\textrm{Br}\big( Z\to\mu e \big)$ and $\textrm{Br}\big( Z\to\tau\mu \big)$, respectively. The graphs of Fig.~\ref{broaderanalysis} share features with those of Fig.~\ref{degenerateetamax}, namely, the regions excluded by current experimental data (top of the graphs) and projected sensitivity from FCC-ee and CEPC (dashed horizontal line). Regions colored in blue correspond to $Y^v\sim\mathcal{O}(1)$, whereas $Y^v\sim\mathcal{O}(10^{-1})$ characterizes purple-colored regions. Regarding the former, note that there is a solid red curve in each of the graphs, which establishes upper limits on the allowed regions, imposed by Eq.~\eqref{pertubativitylimit}, in order to stay within the perturbative regime. Both the blue and the purple regions are constituted by a large set of curves for $\textrm{Br}\big( Z\to\ell_\alpha\ell_\beta \big)$ as functions of the HNL mass $m_N$, which decrease as $m_N$ becomes larger; this behavior is due to the $\eta$ matrix dependence on the HNL mass, which goes as $\eta\propto\frac{1}{m_N^2}$, as dictated by Eq.~\eqref{etainetamax}. Each of the graphs of Fig.~\ref{broaderanalysis} includes a dot-dashed curve which is a plot of the corresponding branching ratio obtained under the assumption $\eta=\eta_\textrm{max}$. Accordingly, such $\eta_\textrm{max}$-assumed branching-ratio curves are no other than those previously displayed in Fig.~\ref{degenerateetamax} and, moreover, they represent an upper constraint on $\textrm{Br}\big( Z\to\ell_\alpha\ell_\beta \big)$ established by non-unitarity.} \textcolor{Black}{Also note that each point of the dot-dashed curve sets, in accordance with non-unitarity constraints, a sort of lower limit on the HNL mass $m_N$ from which some curve in either the blue or the purple region departs.} \textcolor{Black}{In each graph the reader can observe a gap between the plotted regions and the $\eta_\textrm{max}$ curve, which happens because of the condition $\eta(m_N)\big|_{m_N=m_{N_0}}\approx\eta_\textrm{max}$, occurring instead of an exact equality, which is discussed in Appendix~\ref{etaderivationdegenerate}. So, the graphs of Fig.~\ref{broaderanalysis} include restrictions from non-unitarity effects (dot-dashed), from perturbativity (red solid), from current experimental data (solid horizontal), and from expected sensitivity associated to future machines (dashed horizontal). With all these constraints taken into account, the middle and the lower graphs suggest that, in the framework of the ISSM, the branching ratios for the decay processes $Z\to\mu e$ and $Z\to\tau\mu$ would be well beyond the reach of current experimental sensitivity and, moreover, even beyond future sensitivity projected for the FCC-ee and the CEPC. As far as $\textrm{Br}\big( Z\to\tau e \big)$ is concerned, the graph in the upper panel of Fig.~\ref{broaderanalysis} suggests that, while out of the reach of current experimental data, the effects of the ISSM might be attained as long as $Y^\nu\sim\mathcal{O}(1)$. Furthermore, the constraints imposed by non-unitarity and perturbativity signal that the appropriate HNL-mass range to be explored would be $6.9\,\textrm{TeV}\leqslant m_N\leqslant61.7\,\textrm{TeV}$.
}

\subsection{Numerical estimations: non-degenerate HNL masses}
%%%%%%%%%%
%%%%%%%%%%
%%%%%%%%%%
We now present our quantitative analysis of the branching ratios $\textrm{Br}\big( Z\to\ell_\alpha\ell_\beta \big)$, in the general case in which the masses of the HNLs can be non-degenerate. \textcolor{Black}{We start our discussion considering the expression given in Eq.~\eqref{etainY} for the non-unitarity matrix $\eta$. We first point out that, by contrast with the HNL-mass-degenerate scenario, the possibility that a non-degenerate mass spectrum occurs implies, according to this equation, that the non-unitarity matrix $\eta$ shall, in general, depend on the rotation matrix $R$, involved in the expression given for $\hat{Y}$ in Eq.~\eqref{geometricY}.} \textcolor{Black}{Note that our assumption that $M=\Lambda\zeta_M$, established in Eq.~\eqref{MinLambda}, means that the masses of the HNLs go as $m_{N_j}\approx\Lambda (\zeta_M)_{jj}$ and $m_{X_j}\approx\Lambda (\zeta_M)_{jj}$, with $(\zeta_M)_{jj}>0$ and where $j=1,2,3$. With this in mind, notice that $m_{N_j}=T_{j1}\,m_{N_1}$, in which the $T_{j1}$ are positive-valued parameters, with $T_{11}=1$.} \textcolor{Black}{For the sake of practicality, we reduce the number of free parameters by assuming that $T_{31}=T_{21}$, in which case the HNL fields $N_2$, $N_3$, $X_2$, $X_3$ share the same mass, whereas the masses $m_{N_1}$ and $m_{X_1}$, with $m_{N_1}=m_{X_1}$, differ from them. A further reduction of free parameters is attained by taking $m_{N_1}=\Lambda$, so that the matrix $\zeta_M$ gets the expression
\begin{equation}
\zeta_M=
\left(
\begin{array}{ccc}
1 & 0 & 0
\vspace{0.2cm}
\\
0 & T_{21} & 0
\vspace{0.2cm}
\\
0 & 0 & T_{21}
\end{array}
\right).
\label{Mdegenerate}
\end{equation}
Bear in mind that the value $T_{21}=1$ corresponds to the scenario of HNL degenerate masses. As we did in the previous subsection, we use the parametrization given in Eqs.~\eqref{vecetae} to \eqref{vecetatau}, for the vectors $\vec{\eta}_\alpha$, to characterize the $\hat{Y}$ matrix, as shown in Eq.~\eqref{geometricY}. Evidently, this parametrization does not include the $R$ rotation, given as $R=R\big(\theta,\hat{n}\big)$, where $\hat{n}$ is a unitary vector, determined by two angles and which characterizes the rotation axis, whereas $\theta$ is the rotation angle about $\hat{n}$. Aiming at a reduction of the number of free parameters, with the purpose of carrying out quantitative estimations, we consider solely rotations about principal axes, that is, we use $\hat{n}=\hat{\imath},\hat{\jmath},\hat{k}$, so the rotations under consideration are given in terms of only one angle, $\theta$. In particular, we note that the form assumed for $\zeta_M$, Eq.~\eqref{Mdegenerate}, renders the rotations $R_x$, about the $x$ axis, innocuous to the $\eta$ matrix. Now we consider that the different HNL mass spectra comprised by this analysis must include, as a subcase, the scenario of degenerate HNL masses. Recall that the value $T_{21}=1$, which renders $\zeta_M=\textbf{1}_3$, must lead us to the mass-degenerate scenario and therefore reduce the $\eta$ matrix to Eq.~\eqref{etainetamax}. So, in order to make contact with the scenario of HNL mass degeneration, we assume that the expressions of the moduli $|\vec{\eta}_e|$ and $|\vec{\eta}_\mu|$ in terms of both $|\vec{\eta}_\tau|$ and the entries of $\eta_\textrm{max}$, which we discuss in Appendix~\ref{etaderivationdegenerate} and from which Eq.~\eqref{etainetamax} follows, hold.} \textcolor{Black}{At this point, we observe, from Eq.~\eqref{etainY}, that the $\eta$ matrix depends on a number of free parameters:
\begin{enumerate}
%%%
\item The high-energy scale $\Lambda$.
%%%
\item The power $n$, determining the order of the $m_\textrm{D}$ Yukawa-matrix parameters.
%%%
\item The modulus $|\vec{\eta}_\tau|$.
%%%
\item The tuning parameter $x$.
%%%
\item  The angle $\theta$, associated to the rotation matrix $R$.
%%%
\item The $\zeta_M$ parameter $T_{21}$,
%%%
\end{enumerate}
that is, $\eta=\eta(\Lambda, n, x, |\vec{\eta}_\tau|,\theta, T_{21})$. We spare the reader from the cumbersome and intricate explicit expression for $\eta(\Lambda, n, x, |\vec{\eta}_\tau|,\theta, T_{21})$, while we rather explain our \textit{modus operandi} to extract information on the branching ratios $\textrm{Br}\big( Z\to\ell_\alpha\ell_\beta \big)$. During our discussion developed in Appendix~\ref{etaderivationdegenerate}, contextualized in the scenario of HNL mass degeneration, we established that the non-unitarity matrix lies the nearest it can to $\eta_\textrm{max}$ at $x=0.1944$. Bearing this in mind, for our current discussion we set the equation $\eta\big|_{x=0.1944}=\eta_\textrm{max}$ and then we perform a scan over the remaining parameters, except for the high-energy scale $\Lambda$. We then note that for each set of fixed chosen values $\{ n,|\vec{\eta}_\tau|, \beta, T_{21} \}_\textrm{fixed}$ a total of six equations for $\Lambda$ follow from the comparison of the entries of $\eta\big|_{x=0.1944}$ with those of $\eta_\textrm{max}$. Among the resulting $\Lambda$ solutions, we take the largest one to be the starting point to plot a curve of $\textrm{Br}\big( Z\to\ell_\alpha\ell_\beta \big)$ as a function of $\Lambda$, using the corresponding set of fixed values $\{ n,|\vec{\eta}_\tau|, \theta, T_{21} \}_\textrm{fixed}$, along with $x=0.1944$. Regarding the intervals through which the free parameters run, in order to establish a starting $\Lambda$, we use Eqs.~\eqref{nandeta1} and \eqref{nandeta2} for $n$ and $|\vec{\eta}_\tau|$. As far as the rotation angle $\theta$ is concerned, we consider the values $\theta_y,\theta_z=0,\frac{\pi}{4},\frac{\pi}{2},\frac{3\pi}{4}$, with $\theta_y$ and $\theta_z$ denoting the rotation angles about the $y$ axis and the $z$ axis, respectively. Furthermore, we consider the values $T_{21}=0.5, 0.75, 1, 1.3, 2$ for the $\zeta_M$ parameter. }
\\

\textcolor{Black}{We now present quantitative estimations of branching-ratio contributions, in terms of the high-energy scale $\Lambda$, in the scenario of HNL non-degenerate masses. To this end, we have followed the procedure discussed in the previous paragraph. In the preceding subsection, developed around the scenario of HNL-mass degeneration, we observed that the most favorable cLFV 2-body $Z$-decay process is $Z\to\tau e$. Therefore, from here on we focus our attention in this decay. We refer the reader to the graphs displayed in Figs.~\ref{fullnondeggraphs1} and \ref{fullnondeggraphs2},
\begin{figure}[t]
\centering
\includegraphics[width=7cm]{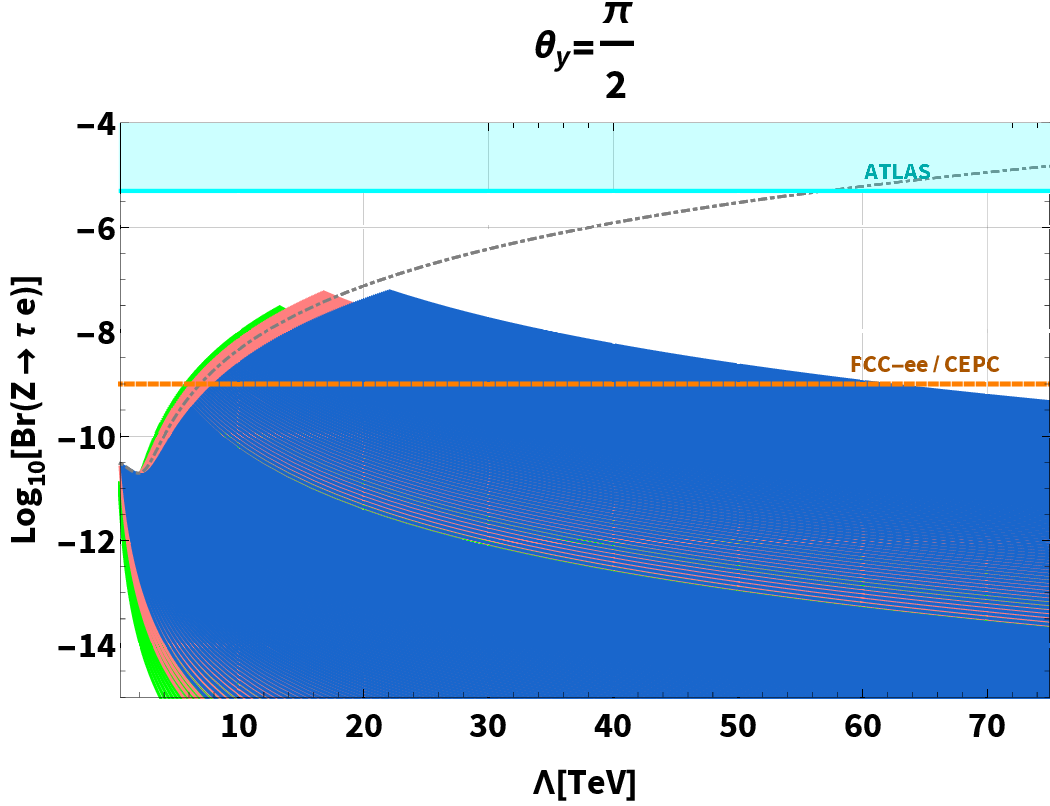}
\vspace{0.2cm} \\
\includegraphics[width=7cm]{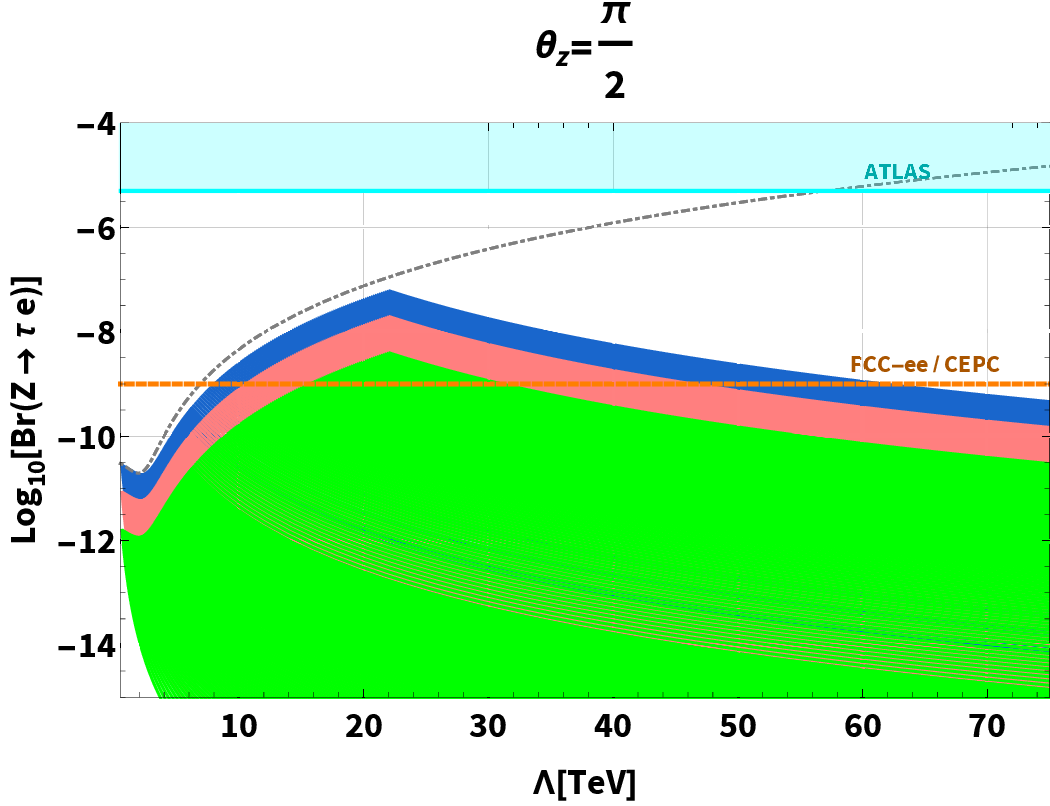}
\vspace{0.2cm} \\
\begin{tikzpicture}
% Cuadro azul
\fill[blue] (0,0) rectangle (0.35,0.35);
\node[anchor=west] at (0.45,0.15)
    {\large$T_{21}=1$};
% Cuadro rojo
\fill[red!50] (3.0,0) rectangle (3.35,0.35);
\node[anchor=west] at (3.45,0.15)
    {\large$T_{21}=1.3$};
% Cuadro verde
\fill[green] (6.0,0) rectangle (6.35,0.35);
\node[anchor=west] at (6.45,0.15)
    {\large$T_{21}=2$};
\end{tikzpicture}
\caption{\label{fullnondeggraphs1} \textcolor{Black}{Allowed regions for $\textrm{Br}\big( Z\to\tau e \big)$, with $1\leqslant T_{21}$, in the $\big( \textrm{Br}(Z\to\tau e),\Lambda \big)$ parameter space. Regions corresponding to the $T_{21}$-parameter choices $T_{21}=1$ (blue), $T_{21}=1.3$ (red), and $T_{21}=2$ (green) are displayed. The branching-ratio values are given in base-10 logarithmic scale. Fixed rotations about either the $y$ axis (upper panel), by an angle $\theta_y=\frac{\pi}{2}$, or about the $z$ axis (lower panel), by an angle $\theta_z=\frac{\pi}{2}$, have been implemented. The regions are limited from above by non-unitarity constraints and validity of the perturbative regime.}} 
\end{figure}
%%%%%%%%%%%%%%%%%%%%
\begin{figure}[t]
\centering
\includegraphics[width=7cm]{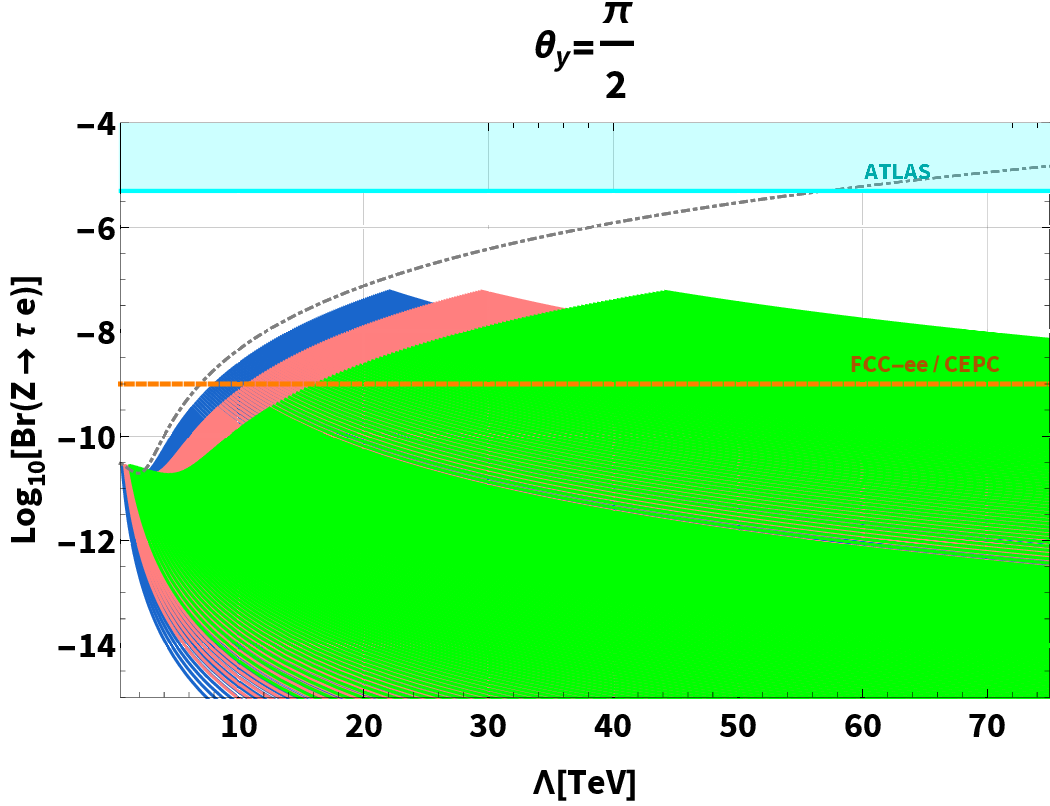}
\vspace{0.2cm} \\
\includegraphics[width=7cm]{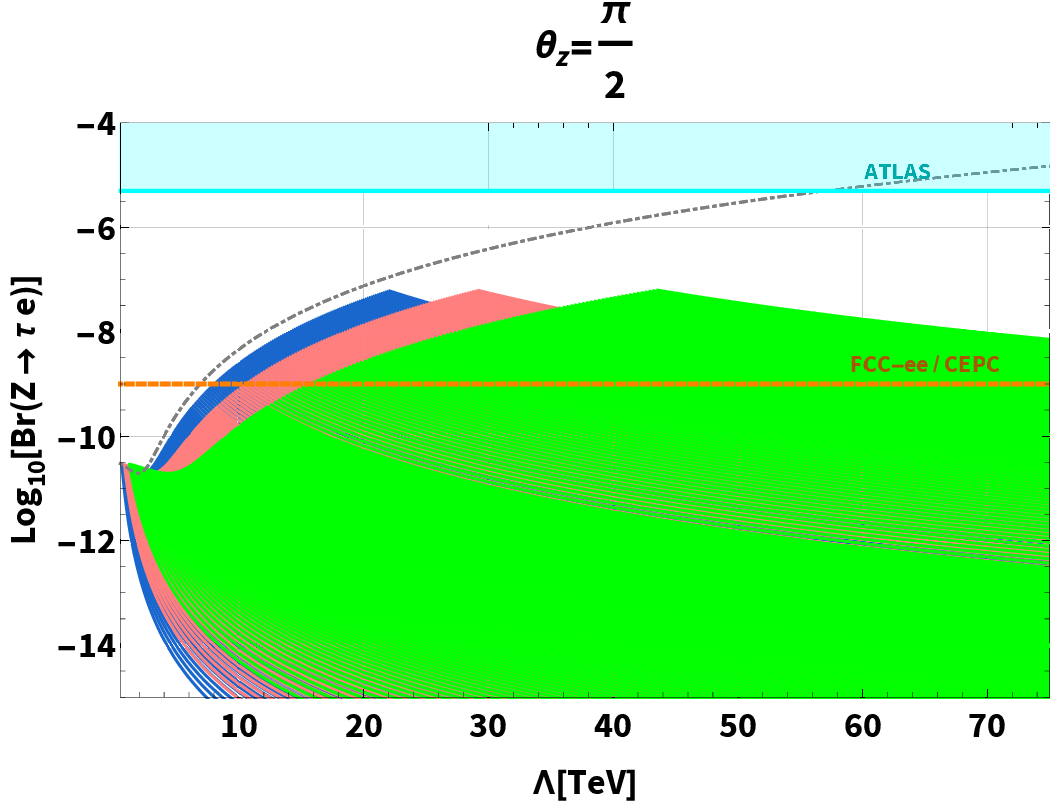}
\vspace{0.2cm} \\
\begin{tikzpicture}
% Cuadro azul
\fill[blue] (0,0) rectangle (0.35,0.35);
\node[anchor=west] at (0.45,0.15)
    {\large$T_{21}=1$};
% Cuadro rojo
\fill[red!50] (3.0,0) rectangle (3.35,0.35);
\node[anchor=west] at (3.45,0.15)
    {\large$T_{21}=0.75$};
% Cuadro verde
\fill[green] (6.0,0) rectangle (6.35,0.35);
\node[anchor=west] at (6.45,0.15)
    {\large$T_{21}=0.5$};
\end{tikzpicture}
\caption{\label{fullnondeggraphs2} \textcolor{Black}{Allowed regions for $\textrm{Br}\big( Z\to\tau e \big)$, with $T_{21}\leqslant1$, in the $\big( \textrm{Br}(Z\to\tau e),\Lambda \big)$ parameter space. Regions corresponding to the $T_{21}$-parameter choices $T_{21}=1$ (blue), $T_{21}=0.75$ (red), and $T_{21}=0.5$ (green) are displayed. The branching-ratio values are given in base-10 logarithmic scale. Fixed rotations about either the $y$ axis (upper panel), by an angle $\theta_y=\frac{\pi}{2}$, or about the $z$ axis (lower panel), by an angle $\theta_z=\frac{\pi}{2}$, have been implemented.  The regions are limited from above by non-unitarity constraints and validity of the perturbative regime.}}
\end{figure}
each of which has been obtained by the reiterative execution of the afore-described procedure. All these graphs show regions within which $\textrm{Br}\big( Z\to\tau e \big)$ is allowed, for different choices of the $T_{21}$ parameter and fixed rotation angle $\theta=\frac{\pi}{2}$, with the rotation implemented either about $\hat{\jmath}$ or $\hat{k}$. In all the graphs, the axes along which the values of $\textrm{Br}\big( Z\to\tau e \big)$ run are given in base-10 logarithmic scale. The allowed regions in these graphs overlap each other and are limited from above by data on non-unitarity effects, characterized by $\eta_\textrm{max}$, and by validity of the perturbative regime, as stated in Eq.~\eqref{pertubativitylimit}. These regions are filled by curves of $\textrm{Br}\big( Z\to\tau e \big)$ as a function of $\Lambda$, which have been obtained by implementing the strategy discussed in the preceding paragraph. The graphs of Figs.~\ref{fullnondeggraphs1} and \ref{fullnondeggraphs2} share a number of elements. For starters, there appears a cyan region at the top of each graph, limited from below by a solid horizontal line and which represents current experimental sensitivity on the branching ratio for the cLFV decay $Z\to\tau e$. Further, a dashed horizontal line is included in the graphs, which has the objective of representing what the projected sensitivity to this decay for the FCC-ee and the CEPC is. Also, a reference plot (dot-dashed curve) corresponding to $\eta=\eta_\textrm{max}$ in the HNL-mass-degenerate scenario, discussed in the previous section, has been included in each graph. As commented before, in order to generate the graphs, we have implemented either a rotation $\theta_y=\frac{\pi}{2}$ about the $y$ axis (upper panels of Figs.~\ref{fullnondeggraphs1} and \ref{fullnondeggraphs2}) or a rotation $\theta_z=\frac{\pi}{2}$ about the $z$ axis (lower panels of Figs.~\ref{fullnondeggraphs1} and \ref{fullnondeggraphs2}). The allowed branching-ratio regions shown in the graphs of Fig.~\ref{fullnondeggraphs1} correspond to choices of the $\zeta_M$-parameter which abide by $T_{21}\geqslant1$. The following values and color scheme have been used: $T_{21}=1$ (blue); $T_{21}=1.3$ (red); and $T_{21}=2$ (green). On the other hand, values $T_{21}\leqslant1$ for the $\zeta_M$ parameters have been considered to carry out the graphs of Fig.~\ref{fullnondeggraphs2}, for which we have used the following color scheme: $T_{21}=1$ in blue; $T_{21}=0.75$ shown in red; and $T_{21}=0.5$ colored in green. From these figures, one can appreciate that the occurrence of different non-degenerate mass spectra for the HNLs comes along with modifications of the allowed regions for the branching ratio. While certain HNL mass spectra would favor expectations for a future measurement, we note that only slight improvements can take place, since perturbativity and data on non-unitarity prevent $\textrm{Br}\big( Z\to\tau e \big)$ from having much larger values. In general, different picks for $T_{21}$ favor certain $\Lambda$ values in the sense that at such values the branching ratio gets maximal. Of course, different rotations also yield different $\textrm{Br}\big( Z\to\tau e \big)$-allowed regions, which we illustrate by means of the graphs displayed in Fig.~\ref{nondegeneratecontrours}.
\begin{figure*}[ht]
\centering
\includegraphics[width=8.5cm]{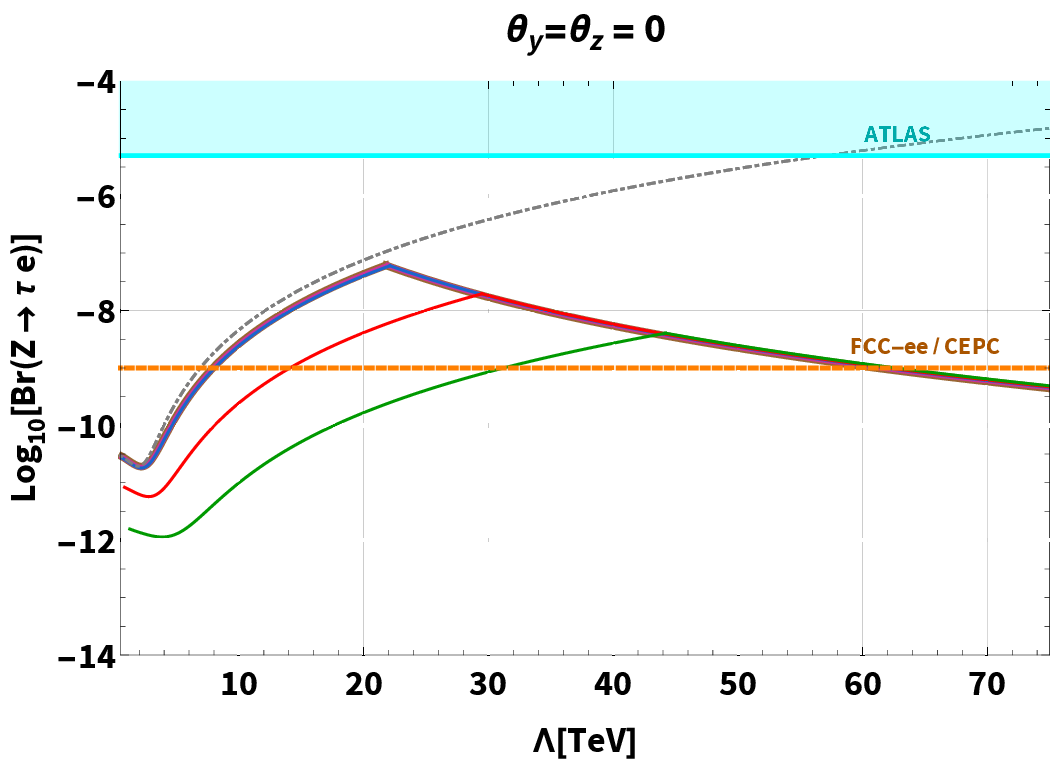}
\hspace{0.3cm}
%%%%%
\includegraphics[width=8.5cm]{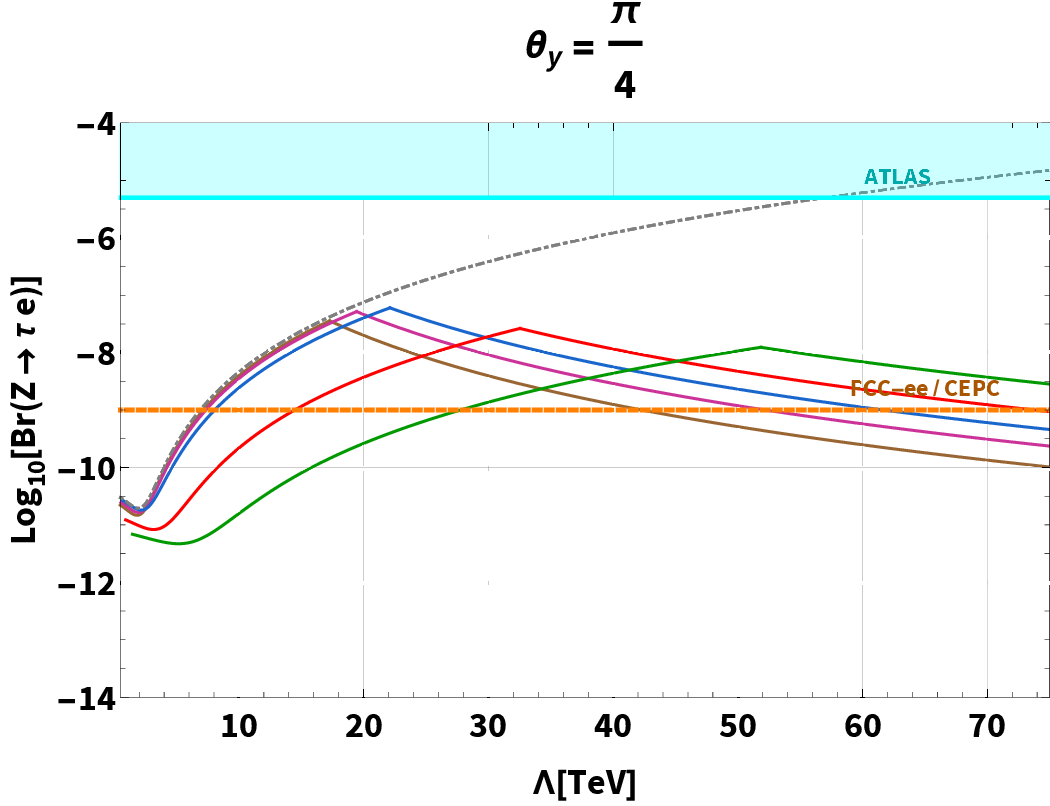}
%%%%%
\vspace{0.7cm} \\
%%%%%
\includegraphics[width=8.5cm]{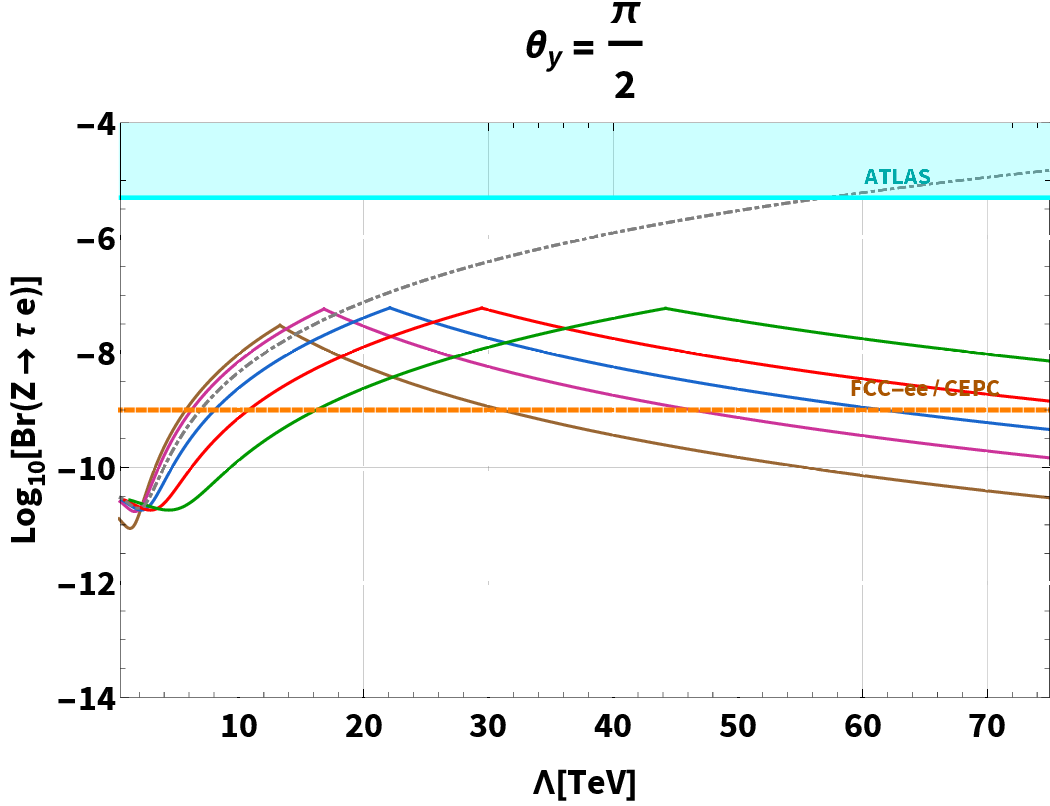}
\hspace{0.3cm}
%%%%%
\includegraphics[width=8.5cm]{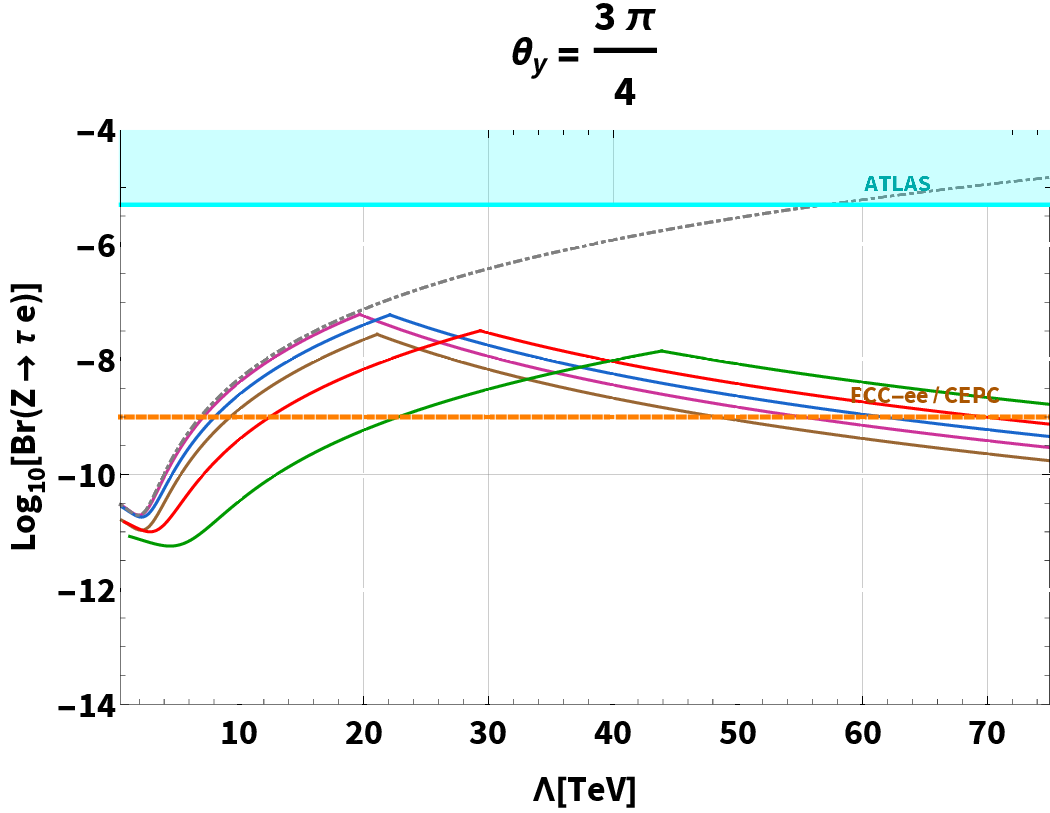}
%%%%%
\vspace{0.4cm} \\
\begin{tikzpicture}
\fill[green!60!black] (-1,0) rectangle (-1.35,0.35);
\node[anchor=west] at (-0.85,0.175)
    {\large$T_{21}=0.5$};
\fill[red] (1.85,0) rectangle (2.20,0.35);
\node[anchor=west] at (2.35,0.175)
    {\large$T_{21}=0.75$};
\fill[blue] (5.30,0) rectangle (5.65,0.35);
\node[anchor=west] at (5.80,0.175)
    {\large$T_{21}=1$};
\fill[magenta!80!black] (7.95,0) rectangle (8.30,0.35);
\node[anchor=west] at (8.45,0.175)
    {\large$T_{21}=1.3$};
\fill[brown] (11.10,0) rectangle (11.45,0.35);
\node[anchor=west] at (11.60,0.175)
    {\large$T_{21}=2$};
\end{tikzpicture}
\caption{\label{nondegeneratecontrours} \textcolor{Black}{Upper bounds, set by non-unitarity constraints and validity of the perturbative regime, for allowed regions in the $\big( \textrm{Br}(Z\to\tau e),\Lambda \big)$ parameter space. Regions corresponding to the parameter values $T_{21}=0.5, 0.75, 1, 1.3, 2$ are displayed, with labeling colors indicated below the graphs. The branching-ratio values are given in base-10 logarithmic scale. Rotations about the $y$ axis, by angles $\theta_y=0$ (upper-left panel), $\theta_y=\frac{\pi}{4}$ (upper-right panel), $\theta_y=\frac{\pi}{2}$ (lower-left panel), and $\theta_y=\frac{3\pi}{4}$ (lower-right panel) have been considered.}}
\end{figure*}
Most elements included in these graphs match those of Figs.~\ref{fullnondeggraphs1} and \ref{fullnondeggraphs2}, which have been already described. Each graph of Fig.~\ref{nondegeneratecontrours} corresponds to a different choice of rotation, for which we have used $\theta_y=0,\frac{\pi}{4},\frac{\pi}{2},\frac{3\pi}{4}$. For clarity purposes, in this case we have only plotted the upper limits on the allowed regions. Moreover, each of the graphs includes five upper-limit curves, associated to the picks $T_{21}=0.5,0.75,1,1.3,2$, with the labeling color scheme indicated in the figure. As can be seen, the afore-commented modification of such regions also takes place.} Again, due to  perturbativity constraints and upper bounds on non-unitarity, no drastic increases in the magnitude of the $\textrm{Br}\big( Z\to\tau e \big)$ contributions are observed. Regarding the order of magnitude of the contributions, we conclude that an optimal context would be characterized by $\textrm{Br}\big( Z\to\tau e \big)\sim10^{-7}$, both in the degenerate and the non-degenerate scenarios of HNL mass. This is to be compared with Refs.~\cite{ARMOT,RHMS}, in which calculations of the branching ratios $\textrm{Br}\big( Z\to\ell_\alpha\ell_\beta \big)$, due to virtual neutral leptons running in one-loop diagrams, in the framework of the ISSM, were carried out. About Ref.~\cite{ARMOT}, the authors perform their analytic calculation and then rely on the Casas-Ibarra parametrization~\cite{CaIb} for their numerical estimations, which is different from the path followed in the present investigation. 
%In relation with this, let us remark that, for our numerical evaluations, the parametrization for the $m_\textrm{D}$ matrix, which essentially is the Yukawa matrix $Y_\nu$ involved in the Yukawa-Dirac term for the right-handed neutrinos $\nu_R$ (see Eq~\eqref{renmassterms}), was neither particular nor unique. Instead, several textures, determined by light-neutrino-mass bounds~\cite{KATRIN}, were considered. 
On the other hand, Ref.~\cite{RHMS} points out how stringent are experimental constraints on cLFV processes of charged leptons in which $\mu-e$ transitions are involved, in comparison with their analogues in which rather $\tau-e$ and $\tau-\mu$ transitions take place, so they propound that a beforehand suppression on $Z\to\mu e$, implemented by a particular sort of parametrization of the Yukawa matrix $Y_\nu$, can be sensibly assumed. Once this has been implemented, the authors of Ref.~\cite{RHMS} focus on the decays $Z\to\tau e$ and $Z\to\tau\mu$, which, favored by the parametrization considered, are enhanced, thus reaching values of order $\sim10^{-7}$, which are \textcolor{Black}{comparable with the branching-ratio contributions presented in the present paper for the decay process $Z\to\tau e$.}

\section{Summary}
\label{summary}
Throughout the present paper, we have described a phenomenological investigation focused in charged-lepton-flavor violation, a physical phenomenon forbidden in the Standard Model, but whose occurrence is guaranteed by the the confirmation of neutrino mass and mixing. While the sole increase of the Standard-Model field content by a set of three $\textrm{SU}(3)_C\otimes\textrm{SU}(2)_L\otimes\textrm{U}(1)_Y$-singlet right-handed neutrino fields, entering through Dirac Yukawa terms, suffices to generate lepton-flavor violation, such an effect is largely suppressed by the Glashow-Iliopoulos-Maiani mechanism, thus calling for a mean to enhance the effect, bringing it closer to current or near-future experimental sensitivity. Among the broad set of low-scale neutrino-mass-generating mechanisms, the so-called inverse seesaw has received much attention. Such a mechanism has been the one taken as the theoretical framework for the present investigation. Among the main processes involving charged-lepton-flavor violation, we consider the $Z$-boson 2-body fermion decays $Z\to\ell_\alpha\ell_\beta$, which bear great relevance in the search for unknown physics, beyond the Standard Model. To be concrete, we have calculated the contributions to $Z\to\ell_\alpha\ell_\beta$, produced by Feynman diagrams involving virtual neutral leptons, both light (the known neutrinos) and heavy, associated to the $\big( 3,3 \big)$ variant of the inverse seesaw mechanism. We have derived an analytic expression for the branching ratio of the generic decay $Z\to\ell_\alpha\ell_\beta$. With general analytic results at hand, we have neglected the masses of charged leptons (external lines) and light neutrinos (internal lines), and then we have \textcolor{Black}{explored, first, the scenario characterized by degenerate mass spectra of heavy-neutral-lepton masses. Afterwards, we have performed a more comprehensive analysis by allowing the occurrence of mass spectra in which the masses of the heavy neutral leptons can differ of each other.} Note that the latter scenario is commonly disregarded in phenomenological investigations, though no \textit{a priori} reason for this, but simplicity, exists. 
%a set of circumstances under which we have been able to derive a simple expression of the branching ratio in which $\textrm{Br}\big( Z\to\ell_\alpha\ell_\beta \big)\propto\big| \eta_{\beta\alpha} \big|^2$, with $\eta$ the $3\times3$ Hermitian matrix characterizing non-unitarity effects in light-lepton mixing, present as a consequence of the existence of heavy neutral leptons. 
Recently upper bounded by the ATLAS Collaboration, at the Large Hadron Collider, the branching ratios associated to these decay processes are expected to be explored with great precision and remarkable improved sensitivity by future in-plan machines, such as the Future Circular Collider, in its electron-positron phase, and by the CEPC. With this in mind, we have carried out numerical estimations of the branching ratios $\textrm{Br}\big( Z\to\ell_\alpha\ell_\beta \big)$ in the context of present and future experimental sensitivity. \textcolor{Black}{Our quantitative results show that the contributions achieved in the scenario of non-degenerate heavy-neutral-lepton mass spectra do not appreciably differ from those emerging under the assumption of mass degeneration, which is caused by both upper bounds on non-unitarity effects and the validity of the perturbative regime.}
%Our quantitative discussion has comprehended two scenarios: (1) the masses of the set of heavy neutral leptons is non-degenerate; and (2) the masses of the heavy-neutral-lepton masses are degenerate, in which case our expression of the branching ratio determined by $\big| \eta_{\beta\alpha} \big|$ is utilized. 
Our estimations and analyses show that, while a measurement of $\textrm{Br}\big( Z\to\ell_\alpha\ell_\beta \big)$ is well beyond current experimental sensitivity, for any $\alpha,\beta=e,\mu,\tau$, future experimental facilities, such as the FCC-ee and the CEPC, would be in position of probing the inverse seesaw mechanism through the decay $Z\to\tau e$. \textcolor{Black}{We arrive at the conclusion that optimal conditions would produce a contribution as large as $\textrm{Br}\big( Z\to\tau e \big)\sim10^{-7}$, which can happen independently of whether the heavy-neutral-lepton mass spectrum is degenerate or not.}

%%%%%%%%%%%%%%%
%%%%%%%%%%%%%%%
%%%%%%%%%%%%%%%
%%%%%%%%%%%%%%%
%%%%%%%%%%%%%%%

\begin{acknowledgments}
The authors acknowledge financial support from SECIHTI (M\'exico). A.G. acknowledges financial support
from SECIHTI `Becas Nacionales para estudios de Posgrado 2023-1' program, with support
number 841506. 
\end{acknowledgments}

%%%%%%%%%%%%%%%
%%%%%%%%%%%%%%%
%%%%%%%%%%%%%%%
%%%%%%%%%%%%%%%
%%%%%%%%%%%%%%%

\appendix

\section{Analytic expressions for the loop contributions}
\label{analyticexpressions}
\textcolor{Black}{We use this Appendix to provide explicit analytic expressions for the functions $f_1$, $f_2$, and $\tilde{f}_2$, which determine the contributions to the branching ratios $\Gamma\big( Z\to\ell_\alpha\ell_\beta \big)$, as shown in Eqs.~\eqref{fAasasum}-\eqref{fA2}. The expressions that we display below correspond to the framework in which the masses of the light neutrinos and the masses of the charged leptons are neglected. We use the short-hand notation
\begin{equation}
B_a^j=B_a\big( 0,m_W^2,m_{\psi_j}^2 \big),
\end{equation}
%%%%%%%%%%
\begin{equation}
C_{abc\ldots}^j=C_{abc\ldots}\big( 0,m_Z^2,0,m_{\psi_j}^2,m_W^2,m_W^2 \big),
\end{equation}
%%%%%%%%%%
\begin{equation}
\hat{C}_{abc\ldots}^{j,k}=C_{abc\ldots}\big( 0,m_Z^2,0,m_W^2,m_{\psi_j}^2,m_{\psi_k}^2 \big),
\end{equation}
to denote Passarino-Veltman functions. Then, we have the expressions
\begin{eqnarray}
&&
f_1(m_{\psi_j})=\frac{1}{4}m_Z^2(2c_\textrm{w}^2-1)(m_{\psi_j}^2
\nonumber \\ && \hspace{0.5cm}
+(D-2)m_W^2)(B_0^j+B_1^j)+m_{\psi_j}^2m_W^2m_Z^2(c_\textrm{w}^2-1)C_0
\nonumber \\ && \hspace{0.5cm}
+c_\textrm{w}^2m_W^2m_Z^4(C_1^j+C_2^j+C_{12}^j)
\nonumber \\ && \hspace{0.5cm}
-\frac{1}{2}m_Z^2\big( (2c_\textrm{w}^2-1)m_{\psi_j}^2+4(D-1)c_\textrm{w}^2m_W^2 \big)C_{00},
\end{eqnarray}
%%%%%%%%%%
\begin{eqnarray}
&&
f_2(m_{\psi_j},m_{\psi_k})=-2(m_{\psi_j}^2m_{\psi_k}^2+2m_W^2m_Z^2)\hat{C}_0^{j,k}
\nonumber \\ && \hspace{0.5cm}
-4m_W^2m_Z^2(\hat{C}_1^{j,k}+\hat{C}_2^{j,k})
\nonumber \\ && \hspace{0.5cm}
+2(D-2)m_W^2\big( (D-2)\hat{C}_{00}^{j,k}-m_Z^2\hat{C}_{12}^{j,k} \big),
\end{eqnarray}
%%%%%%%%%%
\begin{equation}
\tilde{f}_2(m_{\psi_j},m_{\psi_k})=(D-2)(m_W^2\hat{C}_0^{j,k}-\hat{C}_{00}^{j,k})+m_Z^2\hat{C}_{12}^{j,k},
\end{equation}
where $D$ is de dimension of spacetime, in the context of the dimensional regularization method.}

\section{Derivation of the matrix $\eta$ in the HNL-mass degenerate scenario}
\label{etaderivationdegenerate}
\textcolor{Black}{In this Appendix, we discuss the derivation of the expression given in Eq.~\eqref{etainetamax} for the non-unitarity matrix $\eta$. The appeal behind Eq.~\eqref{etainetamax} lies in the small number of parameters required to define $\eta$, as well as its consistency with the bounds on non-unitarity effects given in Refs.~\cite{BFHLMN,GoNo}. By consistency, we mean that this expression for $\eta$ always fulfills the condition $(\eta)_{\alpha\beta}\leqslant(\eta_\textrm{max})_{\alpha\beta}$.}
\\

\textcolor{Black}{We first note that $\textrm{Br}\big( Z\to\ell_\alpha\ell_\beta \big)\equiv\textrm{Br}_{\alpha\beta}$ are determined, in the degenerate HNL-mass scenario, by the $\eta$ entries and by $m_N$, that is, $\textrm{Br}_{\alpha\beta}=\textrm{Br}_{\alpha\beta}(m_N,\eta)$. Moreover, we note that the non-unitarity matrix has expressed as $\eta=\eta(m_N)$ in Eq.~\eqref{etageometricpar}. We are especially interested in an HNL-mass value $m_N=m_{N_0}$ such that 
\begin{equation}
\eta(m_{N_0})=\eta_\textrm{max},
\label{mN0def}
\end{equation}
which we use as a reference mass. In particular, the curves displayed in the graphs of Fig.~\ref{degenerateetamax} should then correspond to $\textrm{Br}_{\alpha\beta}(m_N,\eta(m_{N_0}))$. With this in mind, we use the condition given in Eq.~\eqref{mN0def} and then, by comparing the in-diagonal entries of both sides, we note that the reference mass can be expressed in terms of any of the in-diagonal entries of $\eta$ and $\eta_\textrm{max}$ as
\begin{equation}
m_{N_0}=\frac{v|\vec{\eta}_\alpha|}{\sqrt{2(\eta_\textrm{max})_\alpha}}\times10^n
\end{equation}
with $(\eta_\textrm{max})_\alpha$ denoting the $\alpha$-th entry of the $\eta_\textrm{max}$ diagonal. Since we can choose, we now pick the option $\alpha=\tau$, which, together with Eq.~\eqref{mN0def}, allows us to write the moduli $|\vec{\eta}_e|$ and $|\vec{\eta}_\mu|$, in terms of $|\vec{\eta}_\tau|$, as}\textcolor{Black}{
\begin{equation}
|\vec{\eta}_e|=\frac{(\eta_\textrm{max})_{\tau e}}{(\eta_\textrm{max})_{\tau\tau}}\,|\vec{\eta}_\tau|,
\label{etaeinetatau}
\end{equation}
%%%%%%%%%%
\begin{equation}
|\vec{\eta}_\mu|=\sqrt{\frac{(\eta_\textrm{max})_{\mu\mu}}{(\eta_\textrm{max})_{\tau\tau}}}\,|\vec{\eta}_\tau|.
\label{etamuinetatau}
\end{equation}
These expressions completely determine the in-diagonal entries of the $\eta$ matrix in terms of $|\vec{\eta}_\tau|$. About Eq.~\eqref{etaeinetatau}, in which $|\vec{\eta}_e|$ has been expressed in terms of $|\vec{\eta}_\tau|$, we remark that such an expression was obtained in the limit as $x\to0$, since in this limit the entry $\eta_{\tau e}$ reaches its maximum value, in accordance with Eq.~\eqref{etageometricpar}, which we consider appropriate, as $(\eta_\textrm{max})_{\tau e}$ is enhanced with respect to other off-diagonal $\eta_\textrm{max}$ entries.}
\\

\textcolor{Black}{Implementation of Eqs.~\eqref{etaeinetatau} and \eqref{etamuinetatau} in Eq.~\eqref{etageometricpar} yield Eq.~\eqref{etainetamax}, though a subtlety regarding the $x$ parameter is yet to be discussed. As far as the off-diagonal $\eta$ entries are concerned, they depend on the moduli $|\vec{\eta}_\alpha|$, already determined in terms of $|\vec{\eta}_\tau|$, but they also involve the tuning parameter $x$. We note that an upper limit on $x$, based in the $\eta_\textrm{max}$ restrictions, must be set. The reason is that we have observed, from our quantitative analyses, that $x$ values near 1 could violate the constraint $\eta_{\mu e}\leqslant(\eta_\textrm{max})_{\mu e}$. To this aim, again we use Eq.~\eqref{mN0def} to set relations involving $x$, from which two equations, associated to the $\eta_{\mu e}$ and $\eta_{\tau\mu}$ entries, emerge. We solve them for this parameter, then finding the solutions
\begin{equation}
x_1=\Bigg( \frac{(\eta_\textrm{max})_{\mu e}}{(\eta_\textrm{max})_{\tau e}}\sqrt{ \frac{(\eta_\textrm{max})_{\tau\tau}}{(\eta_\textrm{max})_{\mu\mu}} } \,\Bigg)^\frac{1}{2},
\end{equation}
%%%%%%%%%
%\begin{equation}
%x_2=
%\Bigg(
%\frac{1}{2}
%\Bigg(
%1+\sqrt{1-\frac{4(\eta_\textrm{max})_{\tau\mu}^2}{(\eta_\textrm{max})_{\tau\tau}(\eta_\textrm{max})_{\mu\mu}}}
%\,\Bigg)
%\Bigg)^\frac{1}{2},
%\end{equation}
%%%%%%%%%
\begin{equation}
x_2=
\Bigg(
\frac{1}{2}
\Bigg(
1-\sqrt{1-\frac{4(\eta_\textrm{max})_{\tau\mu}^2}{(\eta_\textrm{max})_{\tau\tau}(\eta_\textrm{max})_{\mu\mu}}}
\,\Bigg)
\Bigg)^\frac{1}{2},
\label{x2solution}
\end{equation}
where we have exclusively considered positive-valued $x$-solutions. Among them, we take the most restrictive one, which turns out to be $x_2=0.1944$. Therefore, using this criterium we set $0<x\leqslant0.1944$. \\
}

\textcolor{Black}{A comment on the consistency of Eq.~\eqref{etainetamax} is opportune. As stated before, one would expect the $\eta$ matrix shown in that equation to fulfill $\eta(m_N)\big|_{m_N=m_{N_0}}=\eta_\textrm{max}$. Nonetheless, from our quantitative analyses, we observe that rather
\begin{equation}
\eta(m_N)\Big|_{m_N=m_{N_0}}\approx\eta_\textrm{max}
\label{consistenteta}
\end{equation}
holds, where $x=x_2=0.1944$ is taken. Such a discrepancy comes from the simplicity of the parametrization based on the $x$ parameter, which simultaneously plays the role of three parameters, namely, the three angles $\theta_{\alpha\beta}$, as shown in Eqs.~\eqref{thetaemu} to \eqref{thetataue}. As we noted in the previous paragraph, a couple of relations generated from the condition given in Eq.~\eqref{mN0def} yield two restrictions on the maximum allowed value of $x$, among which we have used the most restrictive one: $x_2$. However, some of the entries $\eta_{\alpha\beta}$, of the non-unitarity matrix $\eta$, require $x$ values beyond $x_2$, while still lying below $x_1$, to produce their exact reduction to $(\eta_\textrm{max})_{\alpha\beta}$. For a more general $\eta$ parametrization, involving the three angles $\theta_{\alpha\beta}$, a total of three equations relating them could, in principle, be set, then determining restrictions of each of such angles and allowing for the reduction $\eta(m_N)\big|_{m_N=m_{N_0}}=\eta_\textrm{max}$.
}

\end{document}